\documentclass[aps,prd,preprint,superscriptaddress,floatfix,nofootinbib,showkeys,showpacs]{revtex4-1}

\pdfoutput=1

\usepackage{amsmath}
\usepackage{xspace}
\usepackage{listings}
\usepackage{multirow}
\usepackage{lineno}
\usepackage[dvipsnames,svgnames]{xcolor} 
\usepackage{graphicx}
\usepackage{setspace}
 
\usepackage{aas_macros}
\usepackage{natbib} 
\usepackage{hyperref}
\hypersetup{    
  colorlinks      = true,
  linkcolor       = {black},
  linkbordercolor = {white},
  citecolor       = {black},
  citebordercolor = {white},
  urlcolor        = {blue},
  urlbordercolor  = {white},
}
\usepackage[all]{hypcap}
\newcommand{\ie}{i.e.\xspace}
\newcommand{\eg}{e.g.\xspace}
\newcommand{\vs}{vs.\xspace}

\mathchardef\mhyphen="2D
\DeclareMathOperator*{\argmax}{arg\,max}
\newcommand{\vect}[1]{\boldsymbol{#1}}
\newcommand{\roughly}{\ensuremath{ {\sim}\,} }
\newcommand{\gtr}{\ensuremath{ {>}\,} }

\newcommand{\unit}[1]{\ensuremath{\mathrm{\,#1}}\xspace}
\newcommand{\MeV}{\unit{MeV}}
\newcommand{\GeV}{\unit{GeV}}
\newcommand{\TeV}{\unit{TeV}}
\newcommand{\degree}{\ensuremath{{}^{\circ}}\xspace}

\newcommand{\cm}{\unit{cm}}

\newcommand{\kpc}{\unit{kpc}}
\newcommand{\second}{\unit{s}}

\newcommand{\photon}{\unit{ph}}
\newcommand{\sr}{\unit{sr}}

\newcommand{\Lum}{\ensuremath{ L }\xspace}

\newcommand{\secref}[1]{Section~\ref{sec:#1}}

\newcommand{\tabref}[1]{Table~\ref{tab:#1}}

\newcommand{\figref}[1]{Figure~\ref{fig:#1}}

\newcommand{\eqnref}[1]{Equation~\eqref{eqn:#1}}

\newcommand{\Fermi}{\textit{Fermi}\xspace}

\newcommand{\TS}{\ensuremath{\mathrm{TS}}\xspace}

\newcommand{\passsix}{\code{Pass 6}}

\newcommand{\passsevenhyphen}{\code{Pass-7}}
\newcommand{\psevenrep}{\code{P7REP}}
\newcommand{\passeight}{\code{Pass 8}}

\newcommand{\prior}{\ensuremath{\mathcal{P}}\xspace}

\newcommand{\like}{\ensuremath{\mathcal{L}}\xspace} 
\newcommand{\pseudolike}{ {\tilde{\like}} \xspace}   
\newcommand{\loglike}{\ensuremath{\log\like}\xspace}
\newcommand{\given}{\ensuremath{ \,|\, }\xspace}

\newcommand{\data}{ \ensuremath{ \mathcal{D} }\xspace } 

\newcommand{\params}{\ensuremath{\vect{\alpha}}\xspace}
\newcommand{\sig}{\ensuremath{\mu}\xspace}
\newcommand{\bkg}{\ensuremath{\theta}\xspace}
\newcommand{\interest}{\ensuremath{\vect{\sig}}\xspace}
\newcommand{\nuisance}{\ensuremath{\vect{\bkg}}\xspace}

\newcommand{\pvalue}{\textit{p}-value\xspace}

\newcommand{\code}[1]{\lstinline!#1!\xspace}
\newcommand{\ScienceTools}{\code{ScienceTools}}

\newcommand{\pointlike}{\code{pointlike}}

\newcommand{\gtobssim}{\code{gtobssim}}
\newcommand{\HEALPix}{\code{HEALPix}}
\newcommand{\DMFIT}{\code{DMFIT}}
\newcommand{\Pythia}{\code{Pythia}}
\newcommand{\GALPROP}{\code{GALPROP}}

\newcommand{\DM}{\ensuremath{\mathrm{DM}}}
\newcommand{\mDM}{\ensuremath{m_\DM}\xspace}

\newcommand{\sigmav}{\ensuremath{\langle \sigma v \rangle}\xspace}

\newcommand{\PhiPP}{\ensuremath{\Phi_{\rm PP}}\xspace}
\newcommand{\uubar}{\ensuremath{u \bar u}\xspace}
\newcommand{\ddbar}{\ensuremath{d \bar d}\xspace}
\newcommand{\ccbar}{\ensuremath{c \bar c}\xspace}
\newcommand{\ssbar}{\ensuremath{s \bar s}\xspace}
\newcommand{\bbbar}{\ensuremath{b \bar b}\xspace}
\newcommand{\ttbar}{\ensuremath{t \bar t}\xspace}
\newcommand{\ww}{\ensuremath{W^{+}W^{-}}\xspace}
\newcommand{\zz}{\ensuremath{Z^{0}Z^{0}}\xspace}
\newcommand{\gluglu}{\ensuremath{gg}\xspace}
\newcommand{\ee}{\ensuremath{e^{+}e^{-}}\xspace}
\newcommand{\mumu}{\ensuremath{\mu^{+}\mu^{-}}\xspace}
\newcommand{\tautau}{\ensuremath{\tau^{+}\tau^{-}}\xspace}
\newcommand{\relic}{\ensuremath{3\times10^{-26}\cm^{3}\second^{-1}}\xspace}

\newcommand{\Jfactor}{\ensuremath{\mathrm{J\mhyphen factor}}\xspace}
\newcommand{\Jfactors}{\ensuremath{\mathrm{J\mhyphen factors}}\xspace}

\newcommand{\Rmax}{\ensuremath{R_{V_{\rm max}}}\xspace}
\newcommand{\Vmax}{\ensuremath{V_{\rm max}}\xspace}

\newcommand{\rhalf}{\ensuremath{ r_{\rm h} }\xspace}
\newcommand{\Mhalf}{\ensuremath{ M_{\rm h} }\xspace}

\newcommand{\alphas}{\ensuremath{ \alpha_{\rm s} }\xspace}

\newcommand{\NRAND}{300\xspace}
\newcommand{\NRANDTOT}{7500\xspace}
\newcommand{\NSIM}{2000\xspace}
\newcommand{\NSIMTOT}{50000\xspace}

\providecommand\physrep{\ref@jnl{Phys.~Rep.}}%
\providecommand\apjs{\ref@jnl{ApJS}}%
\providecommand{\jcap}{\ref@jnl{JCAP}}%

\begin{document}

\title{Dark Matter Constraints from Observations of 25 Milky Way Satellite Galaxies with the Fermi Large Area Telescope}

\author{M.~Ackermann}
\affiliation{Deutsches Elektronen Synchrotron DESY, D-15738 Zeuthen, Germany}
\author{A.~Albert}
\affiliation{W. W. Hansen Experimental Physics Laboratory, Kavli Institute for Particle Astrophysics and Cosmology, Department of Physics and SLAC National Accelerator Laboratory, Stanford University, Stanford, CA 94305, USA}
\author{B.~Anderson}
\affiliation{Department of Physics, Stockholm University, AlbaNova, SE-106 91 Stockholm, Sweden}
\affiliation{The Oskar Klein Centre for Cosmoparticle Physics, AlbaNova, SE-106 91 Stockholm, Sweden}
\author{L.~Baldini}
\affiliation{Universit\`a  di Pisa and Istituto Nazionale di Fisica Nucleare, Sezione di Pisa I-56127 Pisa, Italy}
\author{J.~Ballet}
\affiliation{Laboratoire AIM, CEA-IRFU/CNRS/Universit\'e Paris Diderot, Service d'Astrophysique, CEA Saclay, 91191 Gif sur Yvette, France}
\author{G.~Barbiellini}
\affiliation{Istituto Nazionale di Fisica Nucleare, Sezione di Trieste, I-34127 Trieste, Italy}
\affiliation{Dipartimento di Fisica, Universit\`a di Trieste, I-34127 Trieste, Italy}
\author{D.~Bastieri}
\affiliation{Istituto Nazionale di Fisica Nucleare, Sezione di Padova, I-35131 Padova, Italy}
\affiliation{Dipartimento di Fisica e Astronomia ``G. Galilei'', Universit\`a di Padova, I-35131 Padova, Italy}
\author{K.~Bechtol}
\affiliation{W. W. Hansen Experimental Physics Laboratory, Kavli Institute for Particle Astrophysics and Cosmology, Department of Physics and SLAC National Accelerator Laboratory, Stanford University, Stanford, CA 94305, USA}
\author{R.~Bellazzini}
\affiliation{Istituto Nazionale di Fisica Nucleare, Sezione di Pisa, I-56127 Pisa, Italy}
\author{E.~Bissaldi}
\affiliation{Istituto Nazionale di Fisica Nucleare, Sezione di Trieste, and Universit\`a di Trieste, I-34127 Trieste, Italy}
% ...

\collaboration{The Fermi-LAT Collaboration}
\noaffiliation

\begin{abstract}

The dwarf spheroidal satellite galaxies of the Milky Way are some of the most dark-matter-dominated objects known.
Due to their proximity, high dark matter content, and lack of astrophysical backgrounds, dwarf spheroidal galaxies are widely considered to be among the most promising targets for the indirect detection of dark matter via $\gamma$ rays. 
Here we report on $\gamma$-ray observations of 25 Milky Way dwarf spheroidal satellite galaxies based on 4 years of \Fermi Large Area Telescope (LAT) data. 
None of the dwarf galaxies are significantly detected in $\gamma$ rays, and we present $\gamma$-ray flux upper limits between 500\MeV and 500\GeV. 
We determine the dark matter content of 18 dwarf spheroidal galaxies from stellar kinematic data and combine LAT observations of 15 dwarf galaxies to constrain the dark matter annihilation cross section.
We set some of the tightest constraints to date on the annihilation of dark matter particles with masses between 2\GeV and 10\TeV into prototypical standard model channels.
We find these results to be robust against systematic uncertainties in the LAT instrument performance, diffuse $\gamma$-ray background modeling, and assumed dark matter density profile.

\end{abstract}


\maketitle
\section{Introduction}

Astrophysical evidence and theoretical arguments suggest that non-baryonic cold dark matter constitutes $\roughly 27\%$ of the contemporary Universe~\citep{Ade:2013zuv}. 
While very little is known about dark matter beyond its gravitational influence, a popular candidate is a weakly interacting massive particle (WIMP)~\citep{Jungman:1995df,Bergstrom:2000pn,Bertone:2004pz}. 
From an initial equilibrium state in the hot, dense early Universe, WIMPs can freeze out with a relic abundance sufficient to constitute much, if not all, of the dark matter. 
Such WIMPs are expected to have a mass in the \GeV to \TeV range and a $s$-wave annihilation cross section usually quoted at $\roughly \relic$ (though, as pointed out by \citet{Steigman:2012nb}, a more accurate calculation yields $\roughly 5\times10^{-26}\cm^3 \second^{-1}$ at masses ${<}\,10\GeV$ and $\roughly 2 \times 10^{-26}\cm^3 \second^{-1}$ at higher masses). 
In regions of high dark matter density, WIMPs may continue to annihilate into standard model particles through processes similar to those that set their relic abundance.

Gamma rays produced from WIMP annihilation, either mono-energetically (from direct annihilation) or with a continuum of energies (through annihilation into intermediate states) may be detectable by the Large Area Telescope (LAT) on board the \textit{Fermi Gamma-ray Space Telescope} (\Fermi)~\citep{Atwood:2009ez}.  
These $\gamma$ rays would be produced preferentially in regions of high dark matter density. 
The LAT has enabled deep searches for dark matter annihilation in the Galactic halo through the study of both continuum and mono-energetic $\gamma$-ray emission~\citep{Hooper:2011ti,Ackermann:2012rg,Abazajian:2012pn,Abdo:2010nc,Ackermann:2012qk,Weniger:2012tx, Ackermann:2013uma}. 
Additionally, robust constraints have been derived from the isotropic $\gamma$-ray background~\citep{Abdo:2010dk, Abazajian:2010sq}, galaxy clusters~\citep{Ackermann:2010rg}, Galactic dark matter substructures~\citep{Zechlin:2011kk,Ackermann:2012nb,Zechlin:2012by}, and the study of nearby dwarf spheroidal galaxies~\citep{Abdo:2010ex,Ackermann:2011wa,GeringerSameth:2011iw,Mazziotta:2012ux,GeringerSameth:2012sr}.

The dwarf spheroidal satellite galaxies of the Milky Way are especially promising targets for the indirect detection of dark matter annihilation due to their large dark matter content, low diffuse Galactic $\gamma$-ray foregrounds, and lack of conventional astrophysical $\gamma$-ray production mechanisms~\citep{Mateo:1998wg,Grcevich:2009gt}. 
An early analysis by \citet{Abdo:2010ex} set constraints on the dark matter annihilation cross section from 11-month LAT observations of 8 individual dwarf spheroidal galaxies. 
In a subsequent effort, \citet{Ackermann:2011wa} presented an improved analysis using 2 years of LAT data, 10 dwarf spheroidal galaxies, and an expanded statistical framework.
The analysis of \citet{Ackermann:2011wa} has two significant advantages over previous analyses.
First, statistical uncertainties in the dark matter content of dwarf galaxies are incorporated as nuisance parameters when fitting for the dark matter annihilation cross section.
Second, observations of the 10 dwarf galaxies are combined into a single joint likelihood analysis for improved sensitivity.
The analysis of \citet{Ackermann:2011wa} probes the canonical thermal relic cross section for dark matter masses $\lesssim 30 \GeV$ annihilating through the \bbbar and \tautau channels, while similar studies by \citet{GeringerSameth:2011iw} and \citet{Mazziotta:2012ux} yield comparable results.

The present analysis expands and improves upon the analysis of \citet{Ackermann:2011wa}.
Specifically, we use a 4-year $\gamma$-ray data sample with an extended energy range from 500\MeV to 500\GeV and improved instrumental calibrations.
We constrain the $\gamma$-ray flux from all 25 known Milky Way dwarf spheroidal satellite galaxies in a manner that is nearly independent of the assumed $\gamma$-ray signal spectrum (\secref{likelihood}).
We use a novel technique to determine the dark matter content of 18 dwarf spheroidal galaxies by deriving prior probabilities for the dark matter distribution from the population of Local Group dwarf galaxies (\secref{darkmatter}). 
This allows us to increase the number of spatially-independent dwarf galaxies in the joint likelihood analysis to 15 (\secref{results}).
We develop a more advanced statistical framework to examine the expected sensitivity of our search.
Additionally, we model the spatial $\gamma$-ray intensity profiles of the dwarf galaxies in a manner that is consistent with their derived dark matter distributions.
We perform an extensive study of systematic effects arising from uncertainties in the instrument performance, diffuse background modeling, and dark matter distribution (\secref{systematics}).
As a final check, we perform a Bayesian analysis that builds on the work of \citet{Mazziotta:2012ux} and yields comparable results (\secref{unfolding}).
No significant $\gamma$-ray emission is found to be coincident with any of the 25 dwarf spheroidal galaxies.
Our combined analysis of 15 dwarf spheroidal galaxies yields no significant detection of dark matter annihilation for particles in the mass range from 2\GeV to 10\TeV annihilating to \ee, \mumu, \tautau, \uubar, \bbbar, or \ww (when kinematically allowed), and we set robust upper limits on the dark matter annihilation cross section.


\section{Data Selection and Preparation}
\label{sec:data}

Since the start of science operations in August of 2008, the LAT has continuously scanned the $\gamma$-ray sky in the energy range from $20\MeV$ to ${>}\,300\GeV$~\citep{Atwood:2009ez}. 
The LAT has unprecedented angular resolution and sensitivity in this energy range, making it an excellent instrument for the discovery of new $\gamma$-ray sources~\citep{Nolan:2011bm}. 
We select a data sample corresponding to events collected during the first four years of LAT operation (2008-08-04 to 2012-08-04). 
We use the \code{P7REP} data set, which utilizes the \passsevenhyphen event reconstruction and classification scheme~\citep{Ackermann:2012kna}, but was reprocessed with improved calibrations for the light yield and asymmetry in the calorimeter crystals~\citep{Bregeon:2013qba}. 
The new calorimeter calibrations improve the in-flight point-spread function (PSF) above $\roughly 3\GeV$ and correct for the small ($\roughly 1\%$ per year), expected degradation in the light yield of the calorimeter crystals measured in flight data. 
Consequently, the absolute energy scale has shifted upward by a few percent in an energy- and time-dependent manner. 
In addition, the re-calibration of the calorimeter light asymmetry leads to a statistical re-shuffling of the events classified as photons.

We select events from the \code{P7REP} \code{CLEAN} class in the energy range from $500\MeV$ to $500\GeV$ and within a 10\degree radius of 25 dwarf spheroidal satellite galaxies~(\figref{fig10}). 
The \code{CLEAN} event class was chosen to minimize particle backgrounds while preserving effective area. 
At high Galactic latitudes in the energy range from $1\GeV$ to $500\GeV$, the particle background contamination in the \code{CLEAN} class is $\roughly 30\%$ of the extragalactic diffuse $\gamma$-ray background~\citep{Ackermann:2012kna}, while between $500\MeV$ and $1\GeV$ the particle background is comparable to systematic uncertainties in the diffuse Galactic $\gamma$-ray emission.
Studies of the extragalactic $\gamma$-ray background at energies greater than $500\GeV$ suggest that at these energies the fractional residual particle background is greater than at lower energies \cite{Ackermann:2013??}.
To reduce $\gamma$-ray contamination from the bright limb of the Earth, we reject events with zenith angles larger than 100\degree and events collected during time periods when the magnitude of rocking angle of the LAT was greater than 52\degree.

We create $14\degree \times 14\degree$ regions-of-interest (ROIs) by binning the LAT data surrounding each of the 25 dwarf galaxies into 0.1\degree pixels and into 24 logarithmically-spaced bins of energy from $500\MeV$ to $500\GeV$.
We model the diffuse background with a structured Galactic $\gamma$-ray emission model ({\it gll\_iem\_v05.fit}) and an isotropic contribution from extragalactic $\gamma$ rays and charged particle contamination ({\it iso\_clean\_v05.txt}).%
\footnote{\url{http://fermi.gsfc.nasa.gov/ssc/data/access/lat/BackgroundModels.html}} 
We build a model of point-like $\gamma$-ray background sources within 15\degree of each dwarf galaxy beginning with the second LAT source catalog (2FGL)~\citep{Nolan:2011bm}. 
We then follow a procedure similar to that of the 2FGL to find additional candidate point-like background sources by creating a residual test statistic map with \pointlike~\citep{Nolan:2011bm}. 
No new sources are found within 1\degree of any dwarf galaxy and the additional candidate sources have a negligible impact on our dwarf galaxy search. 
We use the \code{P7REP\_CLEAN\_V15} instrument response functions (IRFs) corresponding to the LAT data set selected above.
When performing the Bayesian analysis in \secref{unfolding}, we utilize the same LAT data set but follow different data preparation and background modeling procedures, which are described in that section.


\section{Maximum Likelihood Analysis}
\label{sec:likelihood}

Limited $\gamma$-ray statistics and the strong dependence of the LAT performance on event energy and incident direction motivate the use of a maximum likelihood-based analysis to optimize the sensitivity to faint $\gamma$-ray sources. 
We define the standard LAT binned Poisson likelihood,
\begin{equation}
\label{eqn:global}
\like(\interest, \nuisance \given \data ) = \prod_{k} \frac{\lambda_{k}^{n_{k}} e^{-\lambda_{k}}}{n_{k} !},
\end{equation}
as a function of the photon data, $\data$, a set of signal parameters, \interest, and a set of nuisance parameters, \nuisance.
The number of observed counts in each energy and spatial bin, indexed by $k$, depends on the data, $n_{k} = n_{k}(\data)$, while the model-predicted counts depend on the input parameters, $\lambda_{k} = \lambda_{k}(\interest,\nuisance)$. 
This likelihood function encapsulates information about the observed counts, instrument performance, exposure, and background fluxes.
However, this likelihood function is formed ``globally'' (\ie, by tying source spectra across all energy bins simultaneously) and is thus necessarily dependent on the spectral model assumed for the source of interest.
To mitigate this spectral dependence, it is common to independently fit a spectral model in each energy bin, $j$ (\ie, to create a spectral energy distribution for a source)~\cite{Abdo:2009iq}.
This expands the global parameters $\interest$ and $\nuisance$ into sets of independent parameters $\{\interest_j\}$ and $\{\nuisance_j\}$. 
Likewise, the likelihood function in \eqnref{global} can be reformulated as a ``bin-by-bin'' likelihood function,
\begin{equation}\label{eqn:indepebin}
\like(\{\interest_j\}, \{\nuisance_j\} \given \data ) = \prod_{j} \like_j(\interest_j, \nuisance_j \given \data_j )\,. 
\end{equation}
In \eqnref{indepebin} the terms in the product are independent binned Poisson likelihood functions, akin to \eqnref{global} but with the index $k$ only running over spatial bins. 

By analyzing each energy bin separately, we remove the requirement of selecting a global spectrum to span the entire energy range at the expense of introducing additional parameters into the fit.
If the source of interest is bright, it is possible to profile over the spectral parameters of the background sources as nuisance parameters in each energy bin.  
However, for faint or undetected sources it is necessary to devise a slightly modified likelihood scheme where the nuisance parameters in all energy bins are set by the global maximum likelihood estimate,%
\footnote{For brevity we assume that all nuisance parameters are globally fit, but it is straightforward to generalize to the situation where some nuisance parameters are defined independently in each bin.}
\begin{equation}
\label{thetahat}
\hat{\nuisance} = \argmax_{\nuisance} \like(\interest, \nuisance \given \data )\, .
\end{equation}
Fixing the normalizations of the background sources at their globally fit values avoids numerical instabilities resulting from the fine binning in energy and the degeneracy of the diffuse background components at high latitude.
Thus, for the analysis of dwarf galaxies our bin-by-bin likelihood becomes
\begin{equation}\label{eqn:binbybin}
\like(\{\interest_j\}, \hat{\nuisance} \given \data ) = \prod_{j} \like_j(\interest_j, \nuisance_j(\hat{\nuisance}) \given \data_j )\,. 
\end{equation}
The bin-by-bin likelihood is powerful because it makes no assumption about the global signal spectrum.%
\footnote{The second order spectral dependence arising from the choice of a spectral model \textit{within} each bin is found to be ${\sim}2\%$ for our energy bin size.} 
However, it is easy to recreate a global likelihood function to test a given signal spectrum by tying the signal parameters across the energy bins,
\begin{equation}
\label{eqn:global2}
\like(\interest, \hat{\nuisance} \given \data ) =  \prod_{j} \like(\interest_j(\interest), \nuisance_j(\hat{\nuisance}) \given \data_j)\,.
\end{equation}
As a consequence, computing a single bin-by-bin likelihood function allows us to subsequently test many spectral models rapidly.
In practice, the signal parameters derived from the global likelihood function in \eqnref{global2} are found to be equivalent to those derived from \eqnref{global} so long as couplings between the signal and nuisance parameters are small.

For each of the 25 dwarf galaxies listed in~\tabref{dsphs}, we construct a bin-by-bin likelihood function.
Normalizations of the background sources (both point-like and diffuse) are fixed from a global fit over all energy bins.%
\footnote{Including the putative dwarf galaxy source in the global fit has a negligible ($\lesssim 1\%$) impact on the fitted background normalizations.}
The signal spectrum within each bin is modeled by a power law ($\text{d}N/\text{d}E \propto E^{-2}$), and the signal normalization is allowed to vary independently between bins.
In each bin, we set a 95\% confidence level (CL) upper limit on the energy flux from the putative signal source by fixing the parameters of all other bins and finding the value of the energy flux where the log-likelihood has decreased by $2.71/2$ from its maximum (the ``delta-log-likelihood technique'')~\cite{Bartlett:1953,Rolke:2004mj}. 
\figref{fig1} illustrates the detailed bin-by-bin LAT likelihood result for the Draco dwarf spheroidal galaxy.

In \figref{fig2}, we compare the observed bin-by-bin flux upper limits for all 25 dwarf spheroidal galaxies to those derived for \NSIM realistic background-only simulations.
These simulations are generated with the LAT simulation tool \gtobssim using the in-flight pointing history and instrument response of the LAT and including both diffuse and point-like background sources.
We find that the observed limits are consistent with simulations of the null hypothesis. 
We emphasize that these bin-by-bin limits are useful because they make no assumptions about the annihilation channel or mass of the dark matter particle.

While the bin-by-bin likelihood function is essentially independent of spectral assumptions, it does depend on the spatial model of the target source.
The dark matter distributions of some dwarf galaxies may be spatially resolvable by the LAT~\citep{Charbonnier:2011ft}.
Thus, for dwarf galaxies for which stellar kinematic data sets exist, we model the $\gamma$-ray intensity from the line-of-sight integral through the best-fit dark matter distribution as derived in \secref{kinematics}.
Dwarf galaxies that lack stellar kinematic data sets are modeled as point-like $\gamma$-ray sources.
Possible systematic uncertainty associated with the analytic form and spatial extent of the dark matter profile are discussed in \secref{kinematics} and \secref{systematics}.


\section{Dark Matter Annihilation}
\label{sec:darkmatter}
The integrated $\gamma$-ray signal flux at the LAT, $\phi_s$ ($\photon \cm^{-2} \second^{-1}$), expected from dark matter annihilation in a density distribution, $\rho(\vect{r})$, is given by
\begin{equation}
   \phi_s(\Delta\Omega) =
    \underbrace{ \frac{1}{4\pi} \frac{\sigmav}{2m_{\DM}^{2}}\int^{E_{\max}}_{E_{\min}}\frac{\text{d}N_{\gamma}}{\text{d}E_{\gamma}}\text{d}E_{\gamma}}_{\PhiPP}
    \cdot
    \underbrace{\vphantom{\int_{E_{\min}}} \int_{\Delta\Omega}\Big\{\int_{\rm l.o.s.}\rho^{2}(\vect{r})\text{d}l\Big\}\text{d}\Omega '}_{\Jfactor}\,.
    \label{eqn:annihilation}
\end{equation}
Here, the \PhiPP term is strictly dependent on the particle physics properties --- \ie, the thermally-averaged annihilation cross section, \sigmav, the particle mass, $m_\DM$, and the differential $\gamma$-ray yield per annihilation, $\text{d}N_\gamma/\text{d}E_\gamma$, integrated over the experimental energy range.%
\footnote{Strictly speaking, the differential yield per annihilation in Equation (\ref{eqn:annihilation}) is a sum of differential yields into specific final states: ${\text{d}N_\gamma/\text{d}E_\gamma = \sum_f B_f\; \text{d}N^f_\gamma/\text{d}E_\gamma}$, where $B_f$ is the branching fraction into final state $f$. In the following, we make use of Equation (\ref{eqn:annihilation}) in the context of single final states only.}
The \Jfactor is the line-of-sight integral through the dark matter distribution integrated over a solid angle, $\Delta\Omega$. 
Qualitatively, the \Jfactor encapsulates the spatial distribution of the dark matter signal, while \PhiPP sets its spectral character. 
Both \PhiPP and the \Jfactor contribute a normalization factor to the signal flux.

\subsection{Particle Physics Factor}
\label{sec:dmfit}
We generate $\gamma$-ray spectra for dark matter annihilation with the \DMFIT package~\citep{Jeltema:2008hf} as implemented in the LAT \ScienceTools. 
We have upgraded the spectra in \DMFIT from \Pythia~6.4~\citep{Sjostrand:2006za} to \Pythia~8.165~\citep{Sjostrand:2007gs}. 
We generate $\gamma$-ray spectra for the \ee and \mumu channels from analytic formulae, \ie, Equation~(4) in~\citet{Essig:2009jx} (see also~\cite{Beacom:2004pe,Birkedal:2005ep}).  
For the muon channel, we add the contribution from the radiative decay of the muon ($\mu^- \to e^- \, \bar{\nu}_e \, \nu_\mu \, \gamma$) using Equations~(59), (61), and (63) from~\citet{Mardon:2009rc}; this can have a factor-of-two effect on the integrated photon yield for $\mDM \sim 2\GeV$. 
We also extend the range of dark matter masses in \DMFIT to lower ($<10\GeV$) and higher ($>5\TeV$) masses~\citep{Jeltema:2008hf}. 
Thus, the upgraded \DMFIT includes $\gamma$-ray spectra for dark matter annihilation through 12 standard model channels (\ee, \mumu, \tautau, \uubar, \ddbar, \ccbar, \ssbar, \bbbar, \ttbar, \ww, \zz, \gluglu) for 24 dark matter masses between 2\GeV and 10\TeV (when kinematically allowed). 
Annihilation spectra for arbitrary dark matter masses within this range are generated by interpolation.
Appropriate flags are set in \Pythia to generate spectra for dark matter masses below 10\GeV. 
However, we note that we ignore the effect of hadronic resonances, which could be important for some dark matter models with masses less than $10\GeV$.
In addition, the hadronization model used in \Pythia becomes more uncertain at lower energies.
The new spectra are consistent with those calculated in~\cite{Jeltema:2008hf} for dark matter masses between $10\,\GeV$ and $5\,\TeV$. 
Outside of this range we believe that the new spectra are a more accurate representation of the underlying physics since they are derived by 
modeling some of the relevant physical processes, rather than by \textit{ad hoc} extrapolations.
  
\subsection{J-Factors for Dwarf Spheroidal Galaxies} 
\label{sec:kinematics}
The dark matter content of dwarf spheroidal galaxies can be determined through dynamical modeling of their stellar density and velocity dispersion profiles (see a recent review by~\citet{Battaglia:2013wqa}). 
Recent studies have shown that an accurate estimate of the dynamical mass of a dwarf spheroidal galaxy can be derived from measurements of the average stellar velocity dispersion and half-light radius alone~\citep{Walker:2009zp,Wolf:2009tu}. 
The total mass within the approximate half-light radius and likewise the integrated \Jfactor have been found to be fairly insensitive to the assumed dark matter density profile~\citep{Martinez:2009jh,Strigari:2012gn}.
We use the prescription of \citet{Martinez:2013ioa} to calculate \Jfactors for a subset of 18 Milky Way dwarf spheroidal galaxies possessing dynamical mass estimates constrained by stellar kinematic data sets~(\tabref{dsphs})~\cite{Dall'Ora:2006pt, Simon:2007dq, Walker:2008ax, Mateo:2007xh, Koch:2007ye, Simon:2010ek, Munoz:2005be, Willman:2010gy}.

We construct a likelihood function for each individual dwarf spheroidal galaxy from the observed luminosity, half-light radius, \rhalf, and mass within the half-light radius, \Mhalf, together with their associated uncertainties.
We derive priors on the relationships between the luminosity, \Lum, maximum circular velocity, \Vmax, and radius of maximum circular velocity, \Rmax, from the ensemble of Local Group dwarf galaxies~\citep{Strigari:2008ib}.
By implementing this analysis as a two-level Bayesian hierarchical model~\cite{Hobson2010}, we are able to simultaneously constrain both the priors and the resulting likelihoods for the dynamical properties of the dwarf galaxies \cite{Martinez:2013ioa}.
This approach benefits from utilizing all available knowledge of the dark matter distribution in these objects while mitigating any systematic differences between the observed dwarf galaxies and numerical simulations~\cite{Springel:2008cc,Diemand:2006ik}.

While priors on the relationships between, \Lum, \Vmax, and \Rmax can be derived from the data, their form must be chosen \textit{a priori}. 
Motivated by simulations~\cite{Springel:2008cc,Diemand:2006ik}, we assume a prior on the relationship between \Rmax and \Vmax that follows a linear form between $\log(\Rmax)$ and $\log(\Vmax)$ with an intrinsic Gaussian scatter~$\mathcal{N}(\mu,\sigma)$,
\begin{equation}
\mathcal{P}(\log(\Rmax)\,|\,\log(\Vmax)) \approx \mathcal{N}(\alpha_{rv} \log(\Vmax) + \beta_{rv}, \sigma_{rv}).
\end{equation}
Similarly, we assume that the prior in $\log(\Vmax)$ is linearly related to $\log(\Lum)$ with Gaussian scatter,
\begin{equation}
\mathcal{P}(\log(\Vmax)\,|\,\log(\Lum)) \approx \mathcal{N}(\alpha_{vl} \log(\Lum) + \beta_{vl}, \sigma_{vl}).
\end{equation}
Finally, we assume a power-law prior on the galaxy luminosity function~\cite{Koposov:2007ni},
\begin{equation}
\mathcal{P}(\Lum) \approx \Lum^{\alpha_{L}}.
\end{equation}
The 7 free parameters of these priors, $\{ \alpha_{rv}, \beta_{rv}, \sigma_{rv}, \alpha_{vl}, \beta_{vl}, \sigma_{vl}, \alpha_{L}\}$, are derived from the full Local Group dwarf galaxy sample. 
Additionally, each dwarf is described by its measured half-light radius, the mass within this radius, the total luminosity, and the associated errors on these quantities.
We constrain the full set of parameters with a Metropolis nested sampling algorithm with approximately 500,000 points per chain~\citep{Skilling:2004}.

The inner slope of the dark matter density profile in dwarf galaxies remains a topic of debate~\citep{Gilmore:2007fy,Strigari:2010un,Walker:2011zu,Agnello:2012uc}. 
Thus, we repeat the above procedure for two different choices of the dark matter profile: (i) a cuspy Navarro-Frenk-White (NFW) profile~\citep{Navarro:1996gj}, and (ii) a cored Burkert profile~\citep{Burkert:1995yz}:
\begin{align}
 \rho(r) & = \frac{\rho_0 r_s^3}{ r ( r_s + r)^2 } & &  \mathrm{NFW} \label{eq:nfw} \\
 \rho(r) & = \frac{\rho_0 r_s^3}{ (r_s + r) ( r_s^2 + r^2) } & & \mathrm{Burkert} \label{eq:burkert}
\end{align}
After deriving the halo parameters for each form of the dark matter density profile, we perform the line-of-sight integral in \eqnref{annihilation} to calculate the intensity profile and integrated \Jfactor within an angular radius of 0.5\degree ($\Delta\Omega \sim 2.4 \times 10^{-4} \sr$). 
Values of the integrated \Jfactor and spatial extension parameter (defined as the angular size of the scale radius) are included in~\tabref{dsphs}. 
We cross check these results with those derived from an analysis using priors derived from numerical simulations~\cite{Strigari:2007at,Martinez:2009jh,Ackermann:2011wa} and flat, ``non-informative" priors~\cite{Strigari:2008ib, Charbonnier:2011ft}. 
We find that the resulting \Jfactors are consistent within their stated uncertainties. 
We confirm that the integrated \Jfactor within 0.5\degree is fairly insensitive to the choice of dark matter density profile so long as the central value of the slope is less than $1.2$~\cite{Strigari:2007at}. 
Thus, we default to modeling the dark matter distributions within the dwarf galaxies with a NFW profile and examine the impact of this choice in more detail in \secref{systematics}.
\vspace{3.5cm} 

\section{Constraints on Dark Matter}
\label{sec:results}

\subsection{Individual Dwarf Galaxies}
\label{sec:individual}
For each of the 25 dwarf galaxies listed in~\tabref{dsphs}, we use \eqnref{global2} to create global likelihood functions for 6 representative dark matter annihilation channels (\ee, \mumu, \tautau, \uubar, \bbbar, and \ww) and a range of particle masses from 2\GeV to 10\TeV (when kinematically allowed). 
We define a test statistic (\TS) as the difference in the global log-likelihood between the null ($\interest=\interest_0$) and alternative hypotheses ($\interest=\hat{\interest}$)~\cite{Mattox:1996zz,Nolan:2011bm},
\begin{equation}
\TS = -2 \ln \left( \frac {\like(\interest_0, \hat{\nuisance} \given \data)}{\like(\hat{\interest}\,, \hat{\nuisance} \given \data)} \right)\, .
\end{equation}
We note that this definition of \TS differs slightly from that commonly used, since the background parameters are fixed at their best-fit values (\secref{likelihood}).
We find that the \TS distribution derived from simulations is well matched to a $\chi^2/2$ distribution with one bounded degree of freedom, as predicted by the asymptotic theorem of \citet{Chernoff:1954} (\figref{fig5}).
However, the study of random high-latitude blank fields suggests that the significance calculated from asymptotic theorems is an overestimate of the true signal significance (\secref{systematics}).
We find no significant $\gamma$-ray excess coincident with any of the 25 dwarf galaxies for any annihilation channel or mass.%
\footnote{We note that \Jfactors are not required to test the consistency of the $\gamma$-ray data with spectral models of dark matter annihilation.} 

The largest deviation from the null hypothesis, $\TS \sim 7$, occurs when fitting the Sagittarius dwarf spheroidal galaxy. 
Sagittarius is located in a region of intense Galactic foreground emission and is coincident with proposed jet-like structures associated with the \Fermi Bubbles~\citep{Su:2010qj,Su:2012gu}. 
We thus find that systematic changes to the model of the diffuse $\gamma$-ray emission (\secref{systematics}) can lead to \TS changes of $\roughly 5$ in this region.

To derive constraints on the dark matter annihilation cross section, we utilize the kinematically determined \Jfactors derived in \secref{kinematics}.
We select a subset of 18 dwarf galaxies, excluding the 7 galaxies that lack kinematically determined \Jfactors: Bootes~II, Bootes~III, Canis Major, Leo~V, Pisces~II, Sagittarius,\footnote{While some authors have suggested a large dark matter component in the Sagittarius dwarf~\citep{Aharonian:2007km}, tidal stripping of this system leads to complicated stellar kinematics~\citep{Frinchaboy:2012ur}.} and Segue~2.
Statistical uncertainty in the \Jfactor determination is incorporated as a nuisance parameter in the maximum likelihood formulation. 
Thus, the likelihood function for each dwarf galaxy, $i$, is described by,
\begin{equation}
\label{eqn:jlike}
\begin{aligned}
\pseudolike_i(\interest,\params_i \given \data_i) = &\like_i(\interest,\hat{\nuisance_i} \given \data_i) \\
&\times \frac{1}{\ln(10) J_i \sqrt{2 \pi} \sigma_i} e^{-(\log_{10}(J_i)-\overline{\log_{10}(J_i)})^2/2\sigma_i^2} . 
\end{aligned}
\end{equation}
Here, $\like_i$ denotes the individual LAT likelihood function for a single ROI from \eqnref{global2} and the $\params_i$ include both the flux normalizations of background $\gamma$-ray sources (diffuse and point-like) and the associated dwarf galaxy J-factors and statistical uncertainties. 
We fix the background normalizations at their best-fit values and profile over the \Jfactor uncertainties as nuisance parameters to construct a 1-dimensional likelihood function for the dark matter annihilation cross section~\citep{Rolke:2004mj}. 
We note here that the \Jfactor uncertainties must be incorporated as nuisance parameters at the level of the global maximum likelihood fit and not on a bin-by-bin basis. 
Applying \Jfactor uncertainties to each energy bin individually would multi-count these uncertainties when creating the global likelihood function. 

For each of the 18 dwarf galaxies with kinematically determined \Jfactors, we create global likelihood functions incorporating uncertainties on the \Jfactors and derive 95\% CL upper limits on the dark matter annihilation cross section using the delta-log-likelihood technique (Tables~\ref{tab:ee_limits}-\ref{tab:ww_limits}). 
From simulations of the regions surrounding each dwarf galaxy, we confirm the well-documented coverage behavior of this technique, finding that the limit often tends to be somewhat conservative~\citep{Rolke:2004mj}. 
As can be seen from Tables~\ref{tab:tautau_limits}-\ref{tab:bb_limits}, the results for several individual dwarf galaxies (\ie, Draco, Coma Berenices, Ursa Major~II, Segue~1, and Ursa Minor) probe the canonical thermal relic cross section of $\relic$ for low dark matter masses.

\subsection{Combined Analysis}
\label{sec:joint}
Under the assumption that the characteristics of the dark matter particle are shared across the dwarf galaxies, the sensitivity to weak signals can be increased through a combined analysis. 
We create a joint likelihood function from the product of the individual likelihood functions for each dwarf galaxy,
\begin{equation}
\label{eqn:jointlike}
\pseudolike(\interest,\{\params_i\} \given \data ) = \prod_i \pseudolike_i(\interest,\params_i \given \data_i)\,.
\end{equation}
The joint likelihood function ties the dark matter particle characteristics (\ie, cross section, mass, branching ratio, and thus $\gamma$-ray spectrum) across the individual dwarf galaxies. 
From the starting set of 18 dwarf galaxies with kinematically determined \Jfactors, we select a set of non-overlapping dwarfs to ensure statistical independence between observations.
When the ROIs of multiple dwarf galaxies overlap, we retain only the dwarf galaxy with the largest \Jfactor. 
This selection excludes 3 dwarf galaxies: Canes Venatici~I in favor of Canes Venatici~II, Leo~I in favor of Segue~1, and Ursa Major~I in favor of Willman~1. 
Thus, we are left with a subset of 15 dwarf spheroidal galaxies as input to the joint likelihood analysis: Bootes~I, Canes Venatici~II, Carina, Coma Berenices, Draco, Fornax, Hercules, Leo~II, Leo~IV, Sculptor, Segue~1, Sextans, Ursa Major~II, Ursa Minor, and Willman~1.

No significant signal is found for any of the spectra tested in the combined analysis, prompting us to calculate 95\% CL constraints on \sigmav using the delta-log-likelihood procedure described above~(\figref{fig4}).
We repeat the combined analysis on a set of random high-latitude blank fields in the LAT data to calculate the expected sensitivity (\secref{systematics}).
When calculating the expected sensitivity, we randomize the nominal \Jfactors in accord with their measurement uncertainties to form an unconditional ensemble~\cite{Demortier:2003vc,Gross:2011}.
Thus, we note that the positions and widths of the expected sensitivity bands in \figref{fig4} reflect the range of statistical fluctuations expected both from the LAT data and from the stellar kinematics of the dwarf galaxies.
The largest deviation from the null hypothesis occurs for dark matter masses between 10\GeV and 25\GeV annihilating through the \bbbar channel and has $\TS=8.7$. 
The conventional threshold for LAT source discovery is $\TS>25$~\cite{Nolan:2011bm}.
Studies of random blank-sky locations (\secref{systematics}) show that this \TS cannot be naively converted to a significance using asymptotic theorems.
Thus, we derive a global \pvalue directly from random blank high-Galactic-latitude sky positions yielding $p \approx 0.08$.
This global \pvalue includes the correlated trials factor resulting from testing $\roughly 70$ dark matter spectral models. 
As noted by \citet{Hooper:2011ti}, it is difficult to distinguish the $\gamma$-ray spectra of a $\roughly 25\GeV$ dark matter particle annihilating into \bbbar and a lower-mass dark matter particle ($\roughly 5\GeV$) annihilating to \tautau.
However, the \TS quoted above for the \bbbar channel is larger than that found for any of the masses scanned in the \tautau channel. 
Our analysis agrees well with a standard global LAT \ScienceTools analysis using the \code{MINUIT} subroutine \code{MINOS}~\citep{James:1975dr}.

The sensitivity of the combined analysis depends most heavily on the observations of the dwarf galaxies with the largest \Jfactors.
Specifically, the results of the combined analysis are dominated by Coma Berenices, Draco, Segue~1, Ursa Major~II, Ursa Minor, and Willman~1.
Because the combined analysis is dominated by a few dwarf galaxies with high \Jfactors, any statistical fluctuation in the combined results will necessarily be associated to an excess coincident with these dwarf galaxies.
We find that the primary contributors to the deviation of the combined analysis from the null hypothesis are the ultra-faint dwarf galaxies Segue~1, Ursa~Major~II, and Willman~1. 
We find a comparable $\gamma$-ray excess coincident with Hercules and slightly smaller excesses coincident with Sculptor and Canes Venatici~II; however, the relatively small \Jfactors of these dwarf galaxies limit their contribution to the combined analysis.
On the other hand, the lack of any $\gamma$-ray excess coincident with Coma Berenices, Draco, or Ursa Minor decreases the significance of the deviation in the combined analysis.
Thus, we find no significant correlation between the \Jfactor and the $\gamma$-ray \TS of individual dwarf galaxies.
We investigate the impact of removing the ultra-faint dwarf galaxies from the combined analysis at the end of \secref{systematics}.%
\footnote{ The results discussed in \secref{results}, including bin-by-bin likelihood functions, are available in a machine-readable format at: \url{http://www-glast.stanford.edu/pub_data/713/}.}


\section{Systematic Studies}
\label{sec:systematics}
In this section we briefly summarize the process for determining the impact of systematic uncertainties in the LAT instrument performance, the diffuse interstellar $\gamma$-ray emission model, and the assumed dark matter density profile of the dwarf spheroidal galaxies. 
A quantitative summary of these results can be found in~\tabref{systematics}. Additionally, we discuss the analysis of random blank-sky locations to validate the sensitivity and significance as quoted in \secref{results}.
Finally, we consider the impact of removing three ultra-faint dwarf galaxies from the combined analysis.

We follow the ``bracketing IRF'' prescription of~\citet{Ackermann:2012kna} to quantify the impact of uncertainties on the instrument performance.
We re-analyze the LAT data using two sets of IRFs: one meant to maximize the sensitivity of the LAT to weak $\gamma$-ray sources, and the other meant to minimize it, consistent with the uncertainties of the IRFs. 
When creating these bracketing IRFs, we utilize constraints on the systematic uncertainty from~\citet{Ackermann:2012kna}: 10\% uncertainty on the effective area, 15\% uncertainty on the PSF, and a 5\% uncertainty on the energy dispersion. 
The maximized IRFs increase the effective area, decrease the PSF width, and disregard the instrumental energy dispersion. 
Conversely, the minimized IRFs decrease the effective area, increase the PSF width, and include the instrumental energy dispersion. 
For each IRF set we re-run the full LAT data analysis to fit the normalizations of all background sources.
The resulting impact on the combined limits for the dark matter annihilation cross section are shown in~\tabref{systematics}.
Additionally, it should be noted that discrepancies between the IRFs (derived from Monte Carlo simulations) and the true instrument performance could introduce spectral features at the $\roughly 2 \textendash 3\%$ level.
Because the putative dark matter spectra are more sharply peaked than the diffuse background components, spectral features induced by the IRFs could masquerade as small positive signals.
The impact of these effects is not large enough to introduce a globally significant signal, but may be partially responsible for the observed discrepancy between simulations and LAT data.

To quantify the impact of uncertainties on the interstellar emission modeling, we create a set of 8 alternative diffuse models~\cite{dePalma:2013pia}. 
Following the prescription of~\citet{Ackermann:2012aa}, we generate templates for the H{\sc i}, CO, and inverse Compton (IC) emission using the \GALPROP cosmic-ray (CR) propagation and interaction code.%
\footnote{\url{http://galprop.stanford.edu}}
To create 8 reasonably extreme diffuse emission models, we vary the three most influential \GALPROP input parameters as determined by~\citet{Ackermann:2012aa}: the H{\sc i} spin temperature, the CR source distribution, and the CR propagation halo height.
We simultaneously fit the spectral normalizations of the \GALPROP intensity maps, an isotropic component, and all 2FGL sources to 2 years of all-sky LAT data.
Additionally, a log-parabola spectral correction is fit to each model of the diffuse $\gamma$-ray intensity in various Galactocentric annuli.
When performing the fit, we include geometric templates for Loop I~\citep{Casandjian:2009wq} and the \Fermi Bubbles~\citep{Su:2010qj}.
The template for Loop I is based on the model of~\citet{Wolleben:2007pq}, while the template for the \Fermi Bubbles is taken from~\citet{dePalma:2013pia}.
We re-analyze the LAT data surrounding the dwarf spheroidal galaxies with each of the 8 alternative models, and we record the most extreme differences in the upper limits in~\tabref{systematics}.
Because these diffuse models are constructed using a different methodology than the standard LAT interstellar emission model, the resulting uncertainty should be considered as an estimate of the systematic uncertainty due to the interstellar emission modeling process. 
While these 8 models are chosen to be reasonably extreme, we note that they do not span the full systematic uncertainty involved in modeling the interstellar emission. 
We also note that these 8 models do not bracket the official LAT interstellar emission model since the methodologies used to create the models differ. 
We find that variations between the 8 alternative diffuse models have a small impact on the combined limits (\tabref{systematics}); however, they can increase the \TS of the low-mass excess by as much as 50\%.

While stellar kinematic data currently do little to constrain the inner slope of the dark matter density profile in dwarf spheroidal galaxies, the modeling in \secref{kinematics} confirms that the integrated \Jfactor within an angular radius of 0.5\degree is relatively insensitive to the shape of the inner profile~\cite{Strigari:2007at}.
We find that performing the combined analysis assuming a Burkert profile for the dark matter distribution in dwarf galaxies increases the cross section limits by $\roughly 15\%$.
Additionally, we examine the impact of changing the spatial extension of the NFW profiles used to model the dwarf galaxies.
Taking the $\pm 1\sigma$ value of the scale radius (as determined in~\secref{kinematics}) leads to a change of ${<}\,20\%$ in the LAT sensitivity.

We utilize random blank-sky locations as a control sample for the analysis of dwarf spheroidal galaxies. 
(A similar procedure was used by \citet{Mazziotta:2012ux} to analyze the Milky Way dark matter halo.)
We choose blank-sky locations randomly at high Galactic latitude ($|b| > 30\degree$) and far from any 2FGL catalog sources (${>}\,1\degree$ from point-like sources and ${>}\,5\degree$ from spatially extended sources).
We select sets of 25 blank-sky regions separated by $\gtr 7\degree$ to correspond to the 25 dwarf galaxies.
For each blank-sky location, data are selected and binned according to \secref{data}, diffuse background sources and the local 2FGL point-like sources are fit, and the likelihood analysis is performed.
We map each set of random sky locations one-to-one to the dwarf galaxies, and we randomize the nominal \Jfactors according to the uncertainty derived from kinematic measurements.
We form a joint likelihood function from sets of 15 random locations to conform to the 15 dwarf galaxies used in the combined analysis of~\secref{joint}.

In \figref{fig5} we plot the TS values derived by fitting for a source with a $25\GeV$ \bbbar spectrum at each of the \NRANDTOT individual locations, and in \figref{fig4} we show the expected sensitivity from \NRAND combined analyses performed on randomly selected locations.
While the \TS distribution derived from simulations agrees well with asymptotic theorems, it is clear that the TS distribution from random blank fields deviates from this expectation.
The LAT data are known to contain unresolved point-like astrophysical sources which are not included in the background model. 
Additionally, imperfect modeling of the diffuse background emission and percent-level inconsistencies in the determination of the instrument performance may contribute to deviations in the measured \TS distribution.
Each of these systematic effects is a plausible contributor to deviations from the asymptotic expectations, suggesting that the significance of the dark matter  signal hypothesis cannot be naively computed from the \TS using asymptotic theorems.
While the study of random sky locations suffers from limited statistics, increasing the number of randomly selected locations would reduce the independence of each trial.

The combined limits presented in \secref{results} include the ultra-faint dwarf galaxies Segue~1, Ursa Major~II, and Willman~1. 
While Bayesian hierarchical modeling sets rather tight constraints on these dwarf galaxies as members of the dwarf galaxy population, the stellar kinematic data yield larger uncertainties on the \Jfactors of these dwarf galaxies when analyzed individually. 
For example the velocity distribution of Willman~1 appears to be bimodal, and it is not yet clear whether the stars are dynamically bound to this object~\citep{Willman:2010gy}. 
Similarly, the photometry of Ursa Major II shows that this object is elongated; however, the major axis of Ursa Major~II does not point in the direction of the Galactic center and it is unclear if this elongation should be interpreted as a sign of tidal disruption~\citep{Munoz:2010}. 
Removing these three ultra-faint dwarf galaxies has a $\pm 20\%$ impact on the combined limits for soft annihilation spectra (\eg, \bbbar models or \tautau models with $\mDM \lesssim 100\GeV$) and weakens the limits by a factor of $\roughly 2$ for hard annihilation spectra (\eg, \tautau models with $\mDM \gtrsim 500\GeV$); however, it does not significantly impact the constraint on the thermal relic cross section at low dark matter masses (\figref{fig7}).
Moreover, as mentioned in \secref{results}, these ultra-faint dwarf galaxies contribute heavily to the \TS excess observed in the combined analysis.
The removal of all three ultra-faint galaxies reduces the maximum excess from $\TS = 8.7$ to $\TS < 0.5$. 
Conversely, forming a composite analysis from just the three ultra-faint dwarf galaxies yields $\TS \approx 10$, the significance of which must be decreased by an additional trials factor.



\section{Bayesian Analysis}
\label{sec:unfolding}

As a further check of the maximum likelihood analysis, we perform a complementary Bayesian analysis based on the technique of~\citet{Mazziotta:2012ux}.
This approach derives the diffuse background empirically from annuli surrounding the dwarf galaxies and incorporates the instrument response through an iterative Bayesian unfolding~\cite{Abdo:2009iq,Loparco:2009by,Mazziotta:2009rd}.
The Bayesian method discussed here improves upon that of~\citet{Mazziotta:2012ux} by folding the dark matter signal spectrum with the IRFs and incorporating information from all energy bins when reconstructing a posterior probability distribution for the dark matter cross section.
We apply this method to our 4-year data sample and derive upper limits on the dark matter annihilation cross section from a combined analysis of the 15 dwarf galaxies selected in~\secref{joint}.
We find that the results of the Bayesian and maximum likelihood analyses are comparable despite intrinsically different assumptions and methodologies.

We begin by selecting signal and background regions centered on each of the dwarf galaxies.
Signal regions are defined as cones of $0.5\degree$ angular radius,\footnote{The signal region is chosen to correspond to the solid angle used for \Jfactor calculations~(\secref{kinematics}).} while background regions are defined as annuli with inner radii of $5\degree$ and outer radii of $6\degree$.
When defining the background regions, all data within $3\degree$ of 2FGL catalog sources are masked.
Data are binned in energy according to the prescription in~\secref{likelihood}.


Within each energy bin $j$ we denote the number of counts in the signal region as $n_{j}$ and the number of counts in the background region as $m_{j}$.
We assume that the probabilities of measuring $n_{j}$ counts in the signal region and $m_{j}$ counts in the background region are both Poisson distributed with expected values $s_{j} + c b_{j}$ and $b_{j}$, respectively.
Here, $s_{j}$ is the expected number of signal counts (in the signal region), $b_{j}$ is the expected number of background counts (in the background region), while $c$ is the ratio of solid angles between the signal and background regions.\footnote{Point-source masking is accounted for when calculating the ratio of solid angles between the signal and background regions~\cite{Mazziotta:2012ux}.}
The values of $s_{j}$ are tied across the energy bins by the shape of the putative dark matter spectrum and folded with the LAT instrument response.
For each dwarf, the response of the LAT instrument is evaluated from full Monte Carlo simulations of the detector, which take into account the energy dispersion.


Starting from the measured counts in the signal and background regions, we evaluate the Bayesian posterior probability density function (PDF), 
\begin{equation}
\label{eqn:bayes}
\begin{aligned}
p(\sigmav, J, \vect{b}) \propto &~ \like(\sigmav, J, \vect{b} \given \vect{n},\vect{m}) \\
&~ \times \prior(\sigmav) \prior(J) \prior(\vect{b})\,.
\end{aligned}
\end{equation}
Here we indicate with $\vect{n}=\{n_{j}\}$ and $\vect{m}=\{m_{j}\}$ the counts in the signal and background regions and with $\vect{b}=\{b_{j}\}$ the expected counts in the background regions. 
The likelihood $\like(\sigmav, J, \vect{b} \given \vect{n},\vect{m})$ is the product of Poisson probabilities across all energy bins, while $\prior(\sigmav)$, $\prior(J)$, and $\prior(\vect{b})$ are the prior PDFs for the cross section, \Jfactor, and background counts, respectively.
Prior PDFs for the cross section and background counts are assumed to be uniform, while the prior PDF for the \Jfactor is assumed to be log-normal with mean and variance taken from \tabref{dsphs}.
The posterior PDF on $\sigmav$ is evaluated from~\eqnref{bayes}, marginalizing the joint PDF over the nuisance parameters $\vect{b}$ and $J$. 
Upper limits on $\sigmav$ are calculated by integrating the posterior PDF.
We perform a combined analysis of the 15 dwarf galaxies in~\secref{likelihood} using a similar technique, weighting the data associated with each source by its \Jfactor [\eqnref{bayes}].

We validate the performance of the Bayesian analysis by applying it to random blank fields in the high-latitude data chosen in a manner similar to that described in \secref{systematics}.
Similar to the maximum likelihood analysis, we find that the observed cross section limits for low dark matter masses are slightly higher than the median expected from random fields, but are consistent within statistical fluctuations.
The Bayesian analysis presented here is significantly more sensitive to high-mass dark matter models than that presented in~\citet{Mazziotta:2012ux} due to the utilization of all energy bins when testing putative dark matter signal spectra.
However, the upper limits derived from the Bayesian analysis are slightly less constraining than those derived from maximum likelihood analysis.
The differing results may be ascribed to alternative models of the signal (\ie, finite aperture \vs spatial profile) and of the background (\ie, local annulus \vs global template). 
Additionally, differences between the delta-log-likelihood method and the Bayesian method for setting upper limits will lead to more conservative Bayesian limits in the low-counts regime.
Despite differences in background modeling, the treatment of the LAT instrument performance, and the methodology for setting upper limits, the Bayesian and maximum likelihood analyses yield comparable results (\figref{fig6}).


\section{Conclusions}
\label{sec:conclusions}

We have reported on 4-year $\gamma$-ray observations of 25 dwarf spheroidal satellite galaxies of the Milky Way. 
No significant $\gamma$-ray excess was found coincident with any of the dwarf galaxies for any of the spectral models tested. 
We performed a combined analysis of 15 dwarf galaxies under the assumption that the characteristics of the dark matter particle are shared between the dwarfs. 
Again, no globally significant excess was found for any of the spectral models tested. 
We set 95\% CL limits on the thermally averaged dark matter annihilation cross section incorporating statistical uncertainties in the \Jfactors derived from fits to stellar kinematic data. 
These limits constrain the dark matter annihilation cross section to be less than $\relic$ for dark matter particles with a mass less than $10\GeV$ for the \bbbar channel and a mass less than $15\GeV$ for the \uubar and \tautau channels.
Our limits on the dark matter annihilation cross section extend to dark matter masses of $10\TeV$, although Imaging Air Cherenkov Telescopes present more constraining limits at high dark matter masses~(\figref{fig8})~\cite{Aliu:2012ga,Abramowski:2011hc,Aleksic:2011jx}.

The analysis presented here greatly improves our understanding of the statistical and systematic issues involved in the search for dwarf spheroidal galaxies.
Comparing our results to the 2-year maximum likelihood study of 10 dwarf galaxies by~\citet{Ackermann:2011wa}, we find that the current limits are a factor of $\roughly 2$ weaker for soft spectral models and a factor of ${>}\,2$ stronger for hard spectral models. 
These changes can be attributed to updated \Jfactors, the increased photon energy range, the inclusion of more dwarf spheroidal galaxies (specifically Willman~1), and a reclassification of events when moving from \passsix to \psevenrep.
As expected, we find that 4 years of \psevenrep data yield more constraining limits than 2 years of \psevenrep data.

The largest deviation from the null hypothesis occurs for soft $\gamma$-ray spectra and is fit by dark matter in the mass range from 10\GeV to 25\GeV annihilating to \bbbar and has $\TS = 8.7$. 
Control studies of random blank sky locations suggest that the global significance of this excess is $p \approx 0.08$. 
This deviation can also be fit by lower mass dark matter ($\roughly 5\GeV$) annihilating through harder channels (\ie, \tautau).
Much interest in the low-mass WIMP regime has been triggered by $\gamma$-ray studies of the Galactic center~\citep{Hooper:2011ti,Abazajian:2012pn} and by recent direct detection results~\citep{Agnese:2013rvf}.
While suggestive, the analysis of dwarf spheroidal galaxies presented here remains consistent with the null hypothesis within statistical fluctuation.
Constraints on the dark matter annihilation cross section derived here are found to be robust against the systematic uncertainties considered; however, systematics associated with the diffuse modeling are found to more significantly impact the \TS derived for the low-mass excess.

Upcoming improvements to the LAT instrument performance and continued data taking will lead to increased sensitivity for dark matter annihilation in dwarf spheroidal galaxies.
The novel event reconstruction and event selection of \passeight will be especially important as it will both increase the LAT sensitivity to point-like sources and help mitigate systematic effects present in \psevenrep~\cite{Atwood:2013rka}.
Another exciting prospect is the discovery of new dwarf spheroidal galaxies.
Over the past ten years, data from the Sloan Digital Sky Survey (SDSS)~\citep{Abazajian:2008wr} have been used to roughly double the number of known dwarf spheroidal galaxies, while only covering $\roughly 25\%$ of the sky~\citep{Willman:2009dv}.
Thus, the next generations of deep, wide-field photometric surveys (\eg, PanSTARRS~\citep{Kaiser:2002zz}, Southern Sky Survey~\citep{Keller:2007cd}, DES~\citep{Abbott:2005bi}, and LSST~\citep{Ivezic:2008fe}) are expected to greatly increase the number of known Milky Way dwarf spheroidal satellite galaxies~\cite{Tollerud:2008ze}. 
The discovery and characterization of new dwarf spheroidal galaxies could greatly improve the LAT sensitivity to dark matter annihilation in this class of objects.


\section{Acknowledgments}

The \textit{Fermi}-LAT Collaboration acknowledges generous ongoing support from a number of agencies and institutes that have supported both the development and the operation of the LAT as well as scientific data analysis. These include the National Aeronautics and Space Administration and the Department of Energy in the United States, the Commissariat \`a l'Energie Atomique and the Centre National de la Recherche Scientifique / Institut National de Physique Nucl\'eaire et de Physique des Particules in France, the Agenzia Spaziale Italiana and the Istituto Nazionale di Fisica Nucleare in Italy, the Ministry of Education, Culture, Sports, Science and Technology (MEXT), High Energy Accelerator Research Organization (KEK) and Japan Aerospace Exploration Agency (JAXA) in Japan, and the K.~A.~Wallenberg Foundation, the Swedish Research Council and the Swedish National Space Board in Sweden. 

Additional support for science analysis during the operations phase is gratefully acknowledged from the Istituto Nazionale di Astrofisica in Italy and the Centre National d'\'Etudes Spatiales in France.

Support was also provided by the Department of Energy Office of Science Graduate Fellowship Program (DOE SCGF) administered by ORISE-ORAU under Contract No. DE-AC05-06OR23100 and by the Wenner-Gren Foundation.
J.C.~is a Wallenberg Academy Fellow, supported by the Knut and Alice Wallenberg foundation. 
M.R.~received funds from contract FIRB-2012-RBFR12PM1F from the Italian Ministry of Education, University and Research (MIUR).
The authors would like to thank Joakim Edsj{\"o}, Torbj{\"o}rn Sj{\"o}strand, and Peter Skands for helpful conversations concerning Pythia.
The authors acknowledge the use of \HEALPix~\cite{Gorski:2005}.\footnote{\url{http://healpix.sourceforge.net}}

\bibliography{bib}

\pagebreak
\section{Figures \& Tables}
\label{sec:figures}

\newcommand{\capTen}{
Known dwarf spheroidal satellite galaxies of the Milky Way overlaid on a Hammer-Aitoff projection of a 4-year LAT counts map ($E > 1\,\GeV$). 
The 15 dwarf galaxies included in the combined analysis are shown as filled circles, while additional dwarf galaxies are shown as open circles.}
\begin{figure}[h]
  \centering
  \includegraphics{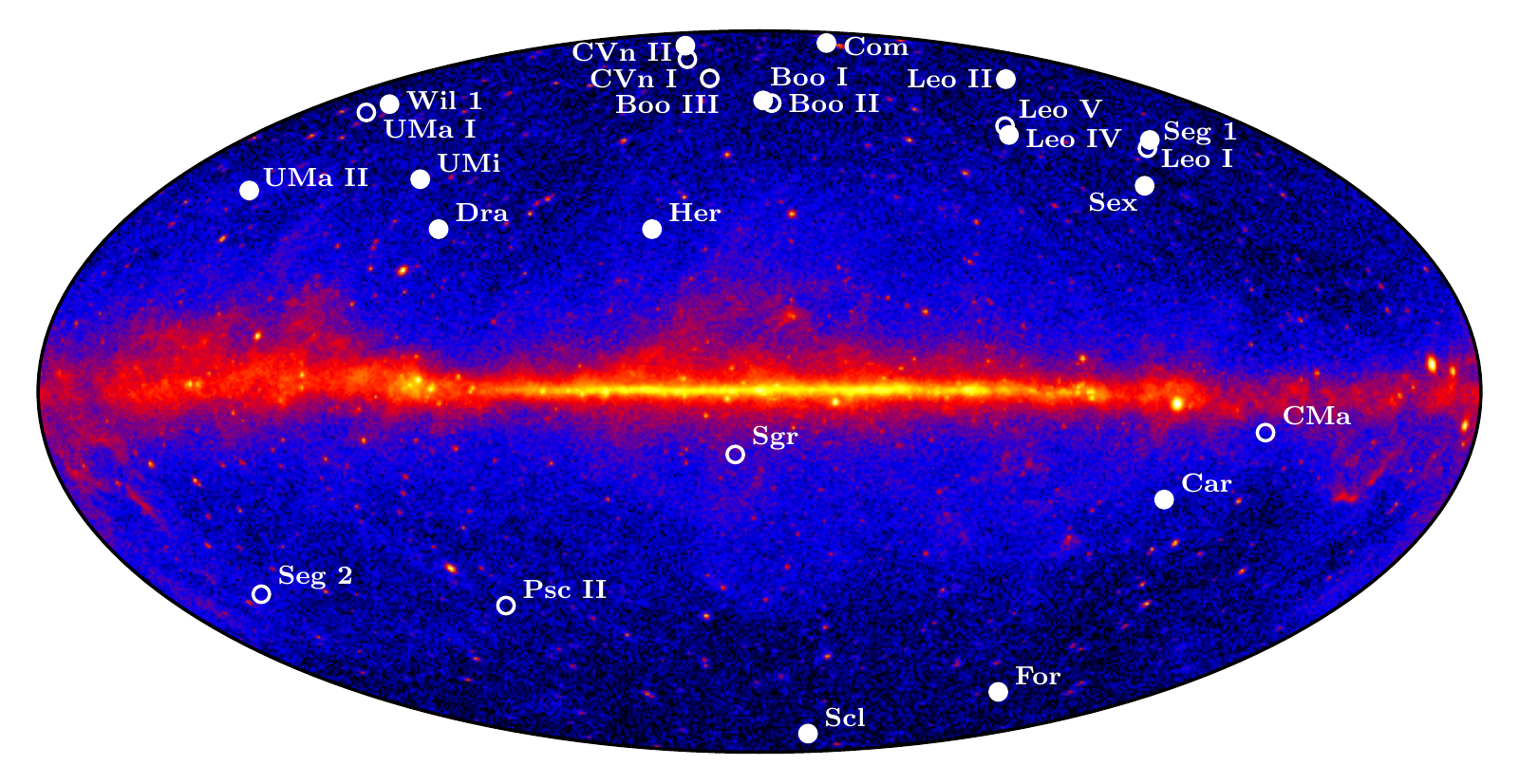}
  \caption{{\capTen}}\label{fig:fig10}
\end{figure}
 
\newcommand{\capOne}{
Histogram of the bin-by-bin LAT likelihood function used to test for a putative $\gamma$-ray source at the position of the Draco dwarf spheroidal galaxy.
The bin-by-bin likelihood is calculated by scanning the integrated energy flux of the putative source within each energy bin (equivalent to scanning in the spectral normalization of the source).
When performing this scan, the flux normalizations of the background sources are fixed to their optimal values as derived from a maximum likelihood fit over the full energy range.
Within each bin, the color scale denotes the variation of the logarithm of the likelihood with respect to the best fit value of the putative source flux (truncated at $-\Delta \loglike > 10$). 
Upper limits on the integrated energy flux are set at 95\% CL within each bin using the delta-log-likelihood technique and are largely independent of the putative source spectrum. }
\begin{figure}[h]
  \centering
  \includegraphics{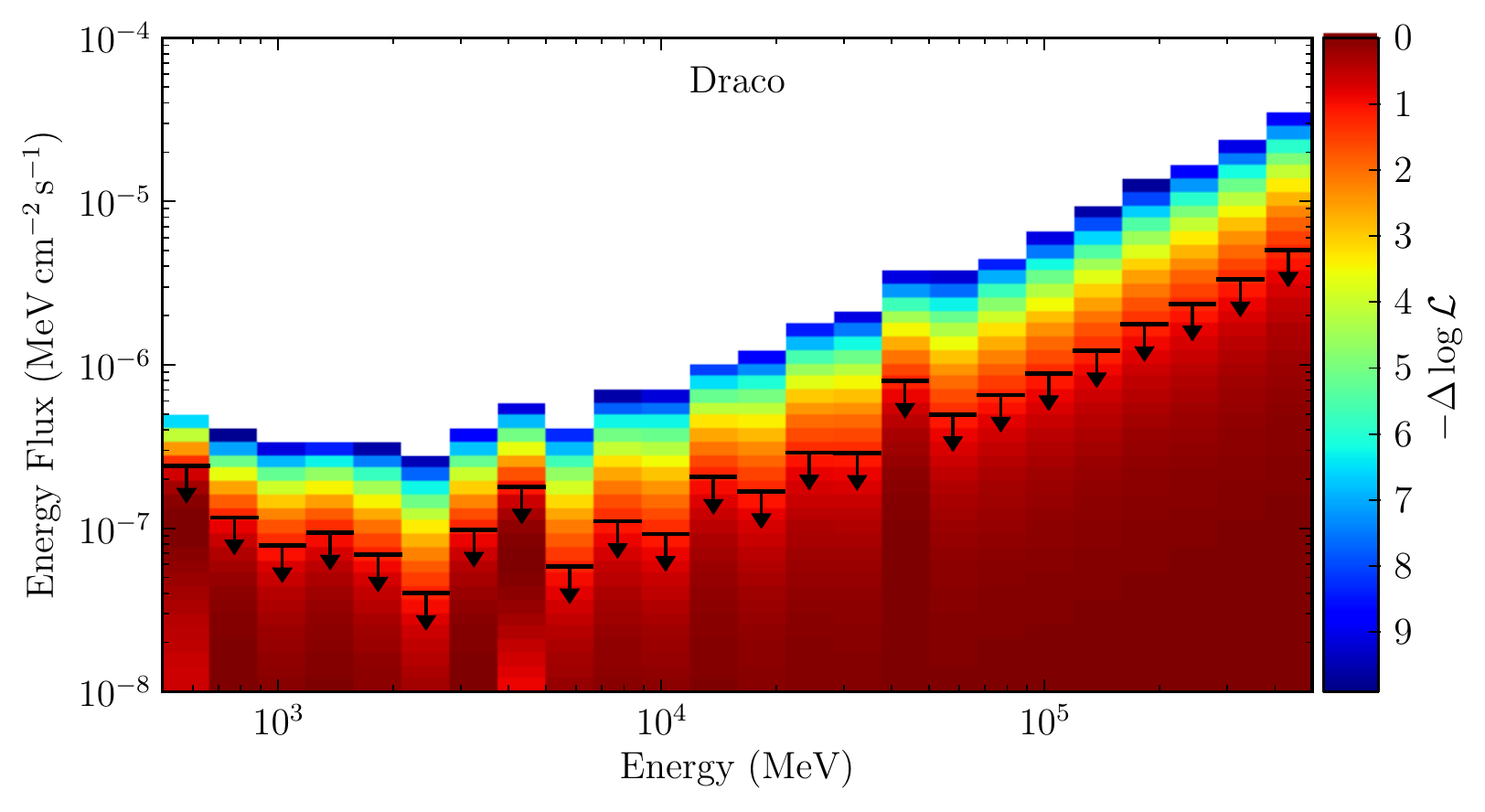}
  \caption{{\capOne}}\label{fig:fig1}
\end{figure}
 
\newcommand{\capTwo}{
Bin-by-bin integrated energy-flux upper limits and expected sensitivities at 95\% CL for each dwarf spheroidal galaxy. 
The expected sensitivities are calculated from \NSIM realistic Monte Carlo simulations of the null hypothesis. 
The median sensitivity is shown by the dashed black line while the 68\% and 95\% containment regions are indicated by the green and yellow bands, respectively.}
\begin{turnpage}
\begin{figure}[h]
  \centering
  \includegraphics{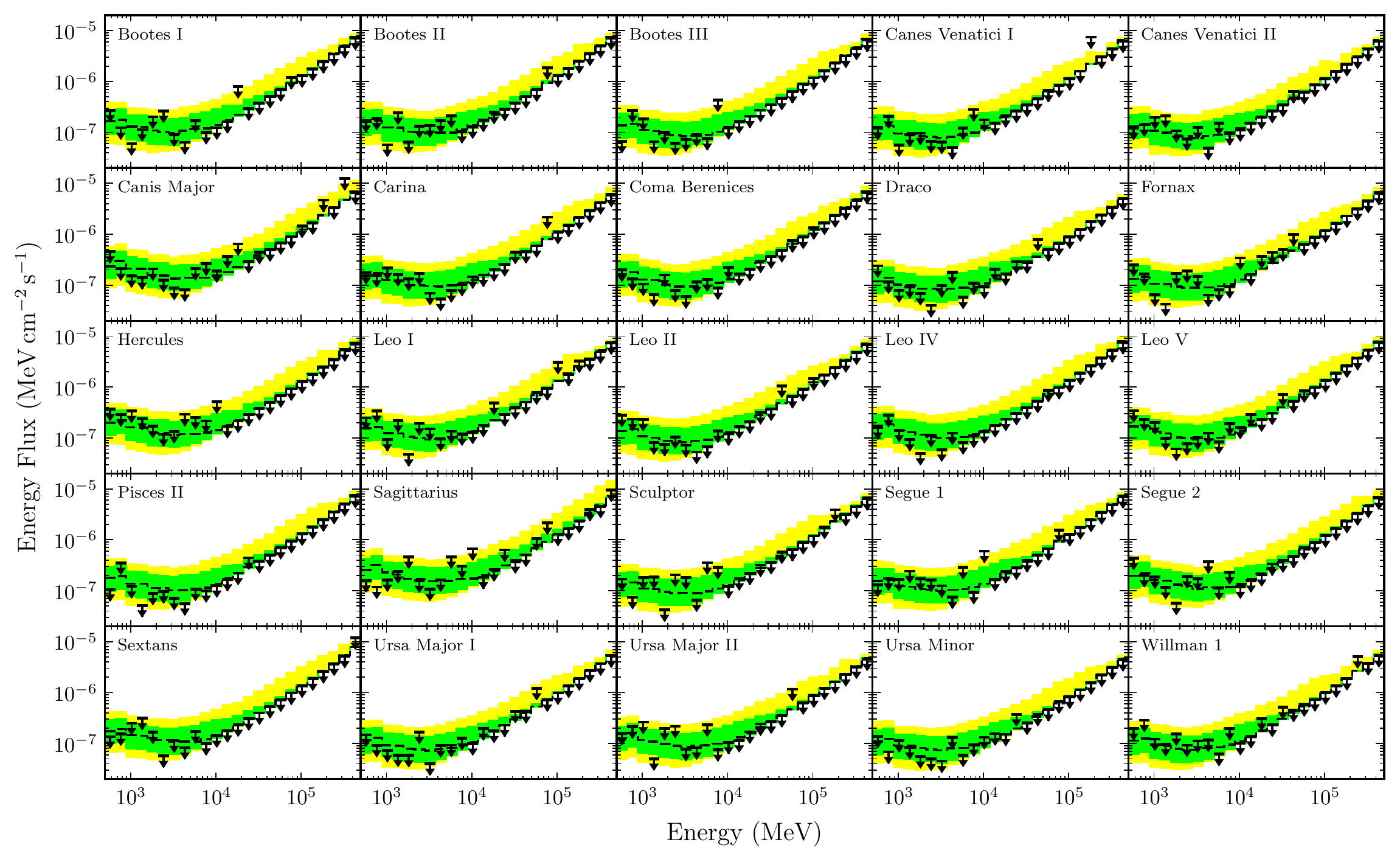}
  \caption{{\capTwo}}\label{fig:fig2}
\end{figure}
\end{turnpage}
 
\newcommand{\capFive}{
Distribution of \TS values from individual fits of a 25\GeV \bbbar annihilation spectrum to the null hypothesis generated from \NSIMTOT realistic Monte Carlo simulations and \NRANDTOT random blank-sky locations at high latitude in the LAT data. 
The distribution of \TS values derived from simulations of individual ROIs is well matched to the expectations from the asymptotic theorem of \citet{Chernoff:1954}, while the distribution derived from random sky positions shows an excess at large \TS values.}
\begin{figure}[h]
  \centering
  \includegraphics{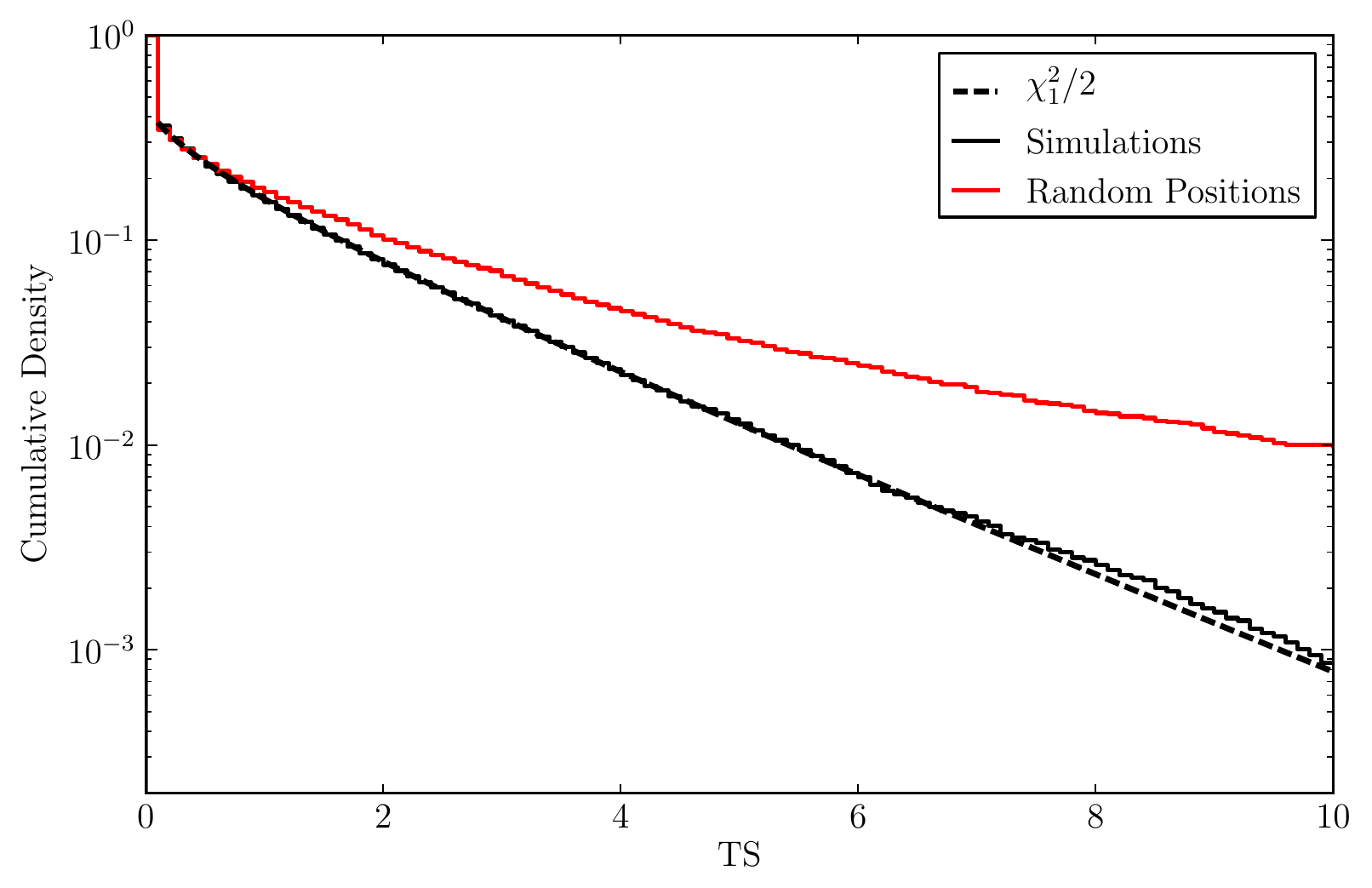}
  \caption{{\capFive}}\label{fig:fig5}
\end{figure}
 
\newcommand{\capFour}{ 
Constraints on the dark matter annihilation cross section at 95\% CL derived from a combined analysis of 15 dwarf spheroidal galaxies assuming an NFW dark matter distribution (solid line). 
In each panel bands represent the expected sensitivity as calculated by repeating the combined analysis on \NRAND randomly-selected sets of blank fields at high Galactic latitudes in the LAT data.
The dashed line shows the median expected sensitivity while the bands represent the 68\% and 95\% quantiles.
For each set of random locations, nominal \Jfactors are randomized in accord with their measurement uncertainties.
Thus, the positions and widths of the expected sensitivity bands reflect the range of statistical fluctuations expected both from the LAT data and from the stellar kinematics of the dwarf galaxies.
The most significant excess in the observed limits occurs for the \bbbar channel between 10\GeV and 25\GeV with $\TS=8.7$ (global \pvalue of $p \approx 0.08$).}
\begin{figure}[h]
  \centering
  \includegraphics{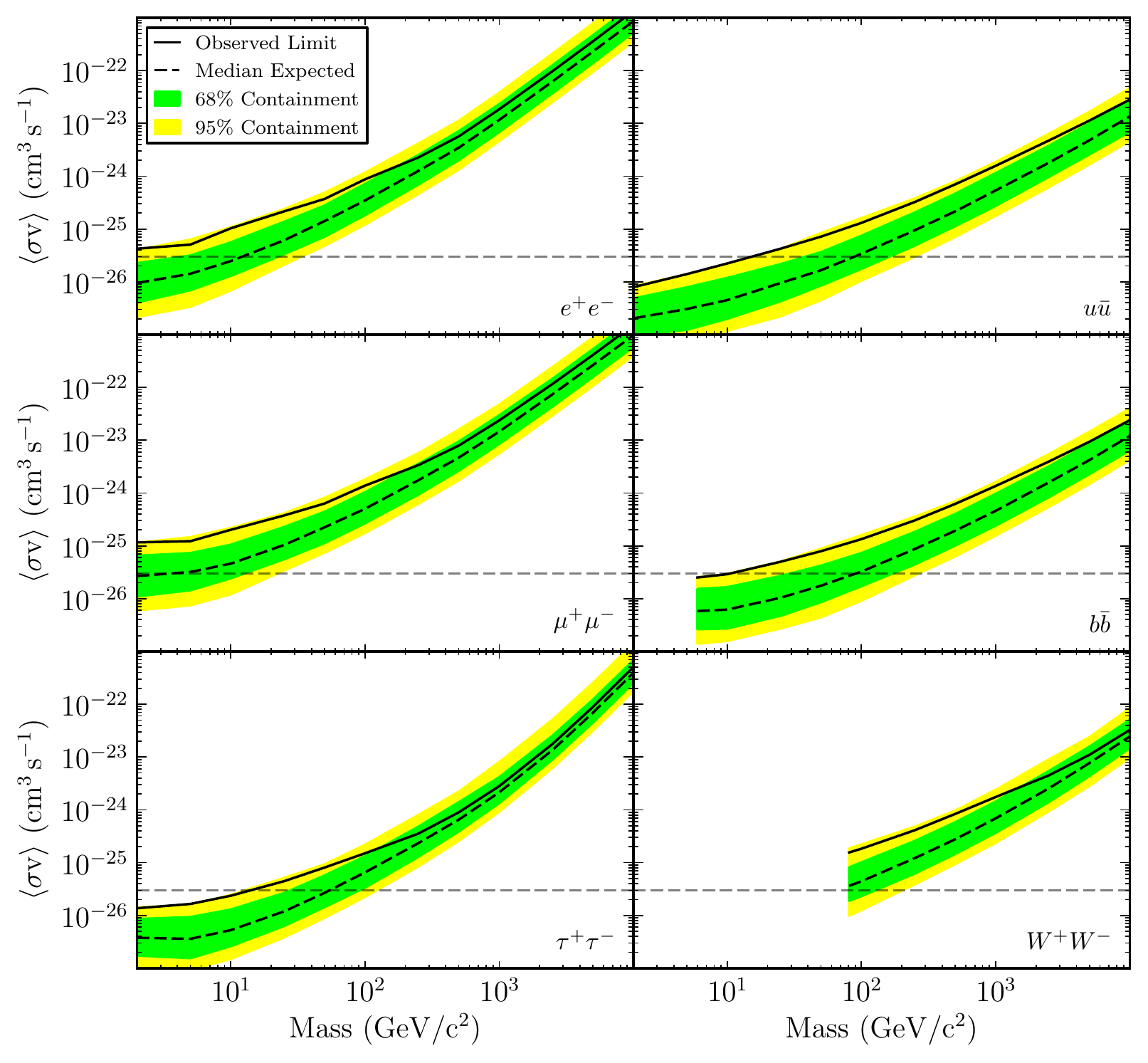}
  \caption{{\capFour}}\label{fig:fig4}
\end{figure}

\newcommand{\capSeven}{ 
Constraints on the dark matter annihilation cross section (\tautau channel) at 95\% CL derived from a combined analysis excluding three ultra-faint dwarf galaxies: Segue~1, Ursa Major~II, and Willman~1 (solid line). 
The expected sensitivity is also calculated excluding these three ultra-faint dwarf galaxies and is represented in the same manner as in \figref{fig4}.
}
\begin{figure}[h]
  \centering
  \includegraphics{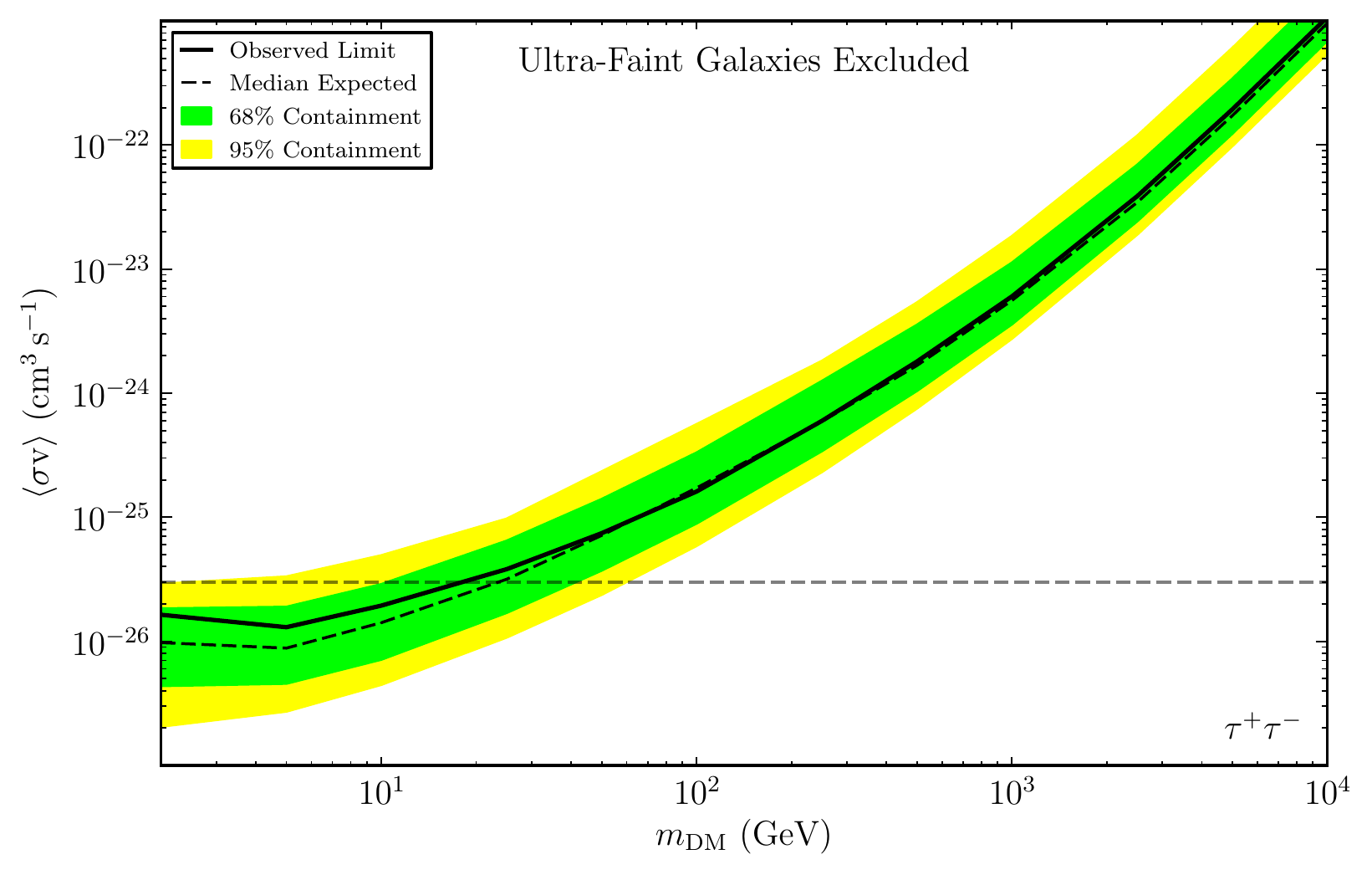}
  \caption{{\capSeven}}\label{fig:fig7}
\end{figure}
 
\newcommand{\capSix}{ 
Comparison of constraints on the dark matter annihilation cross section (\tautau channel) derived from the combined maximum likelihood and the combined Bayesian analyses of 15 dwarf spheroidal galaxies. 
The expected sensitivity for the maximum likelihood analysis is represented similarly to \figref{fig4}.
The observed Bayesian limits are consistent with the expected Bayesian sensitivity bands (not shown), which are likewise higher than those of the maximum likelihood analysis.
}
\begin{figure}[h]
  \centering
  \includegraphics{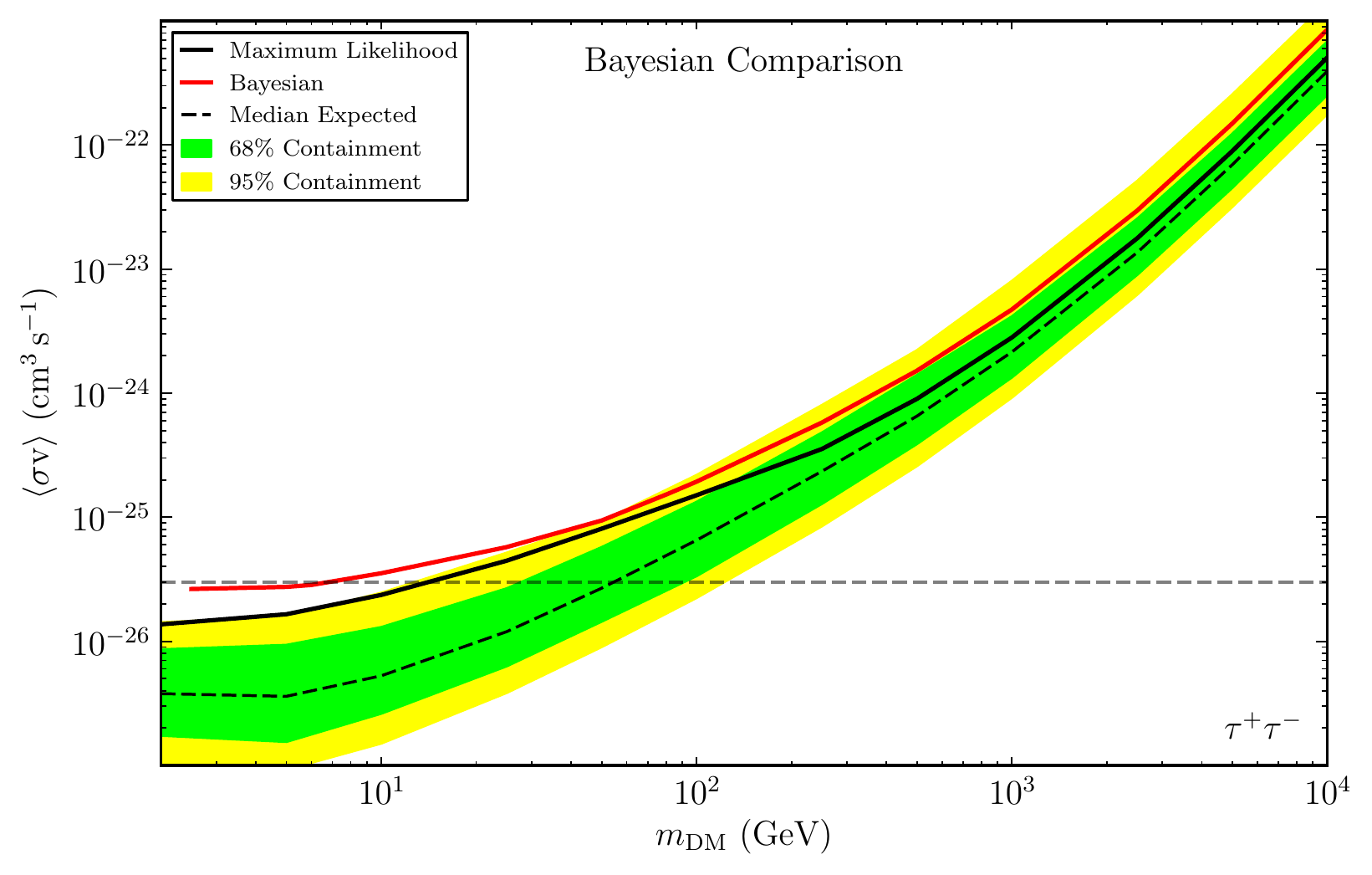}
  \caption{{\capSix}}\label{fig:fig6}
\end{figure}

\newcommand{\capEight}{ 
Comparison of constraints on the dark matter annihilation cross section (\bbbar channel) derived from the LAT combined analysis of 15 dwarf galaxies (assuming an NFW profile), 48-hour observations of Segue~1 by VERITAS (assuming an Einasto profile)~\cite{Aliu:2012ga}, and 112-hour observations of the Galactic center by H.E.S.S. (assuming an Einasto profile)~\cite{Abramowski:2011hc}. 
In the interest of a direct comparison, we also show the LAT constraints derived for Segue~1 alone assuming an Einasto dark matter profile consistent with that used by VERITAS~\cite{Aliu:2012ga}.
For this rescaling, the \Jfactor of Segue~1 is calculated over the LAT solid angle of $\Delta\Omega \sim 2.4 \times 10^{-4} \sr$ and yields a rescaled value of $1.7 \times 10^{19} \GeV^2 \cm^{-5} \sr$ (uncertainties on the \Jfactor are neglected for comparison with VERITAS).
}
\begin{figure}[h]
  \centering
  \includegraphics{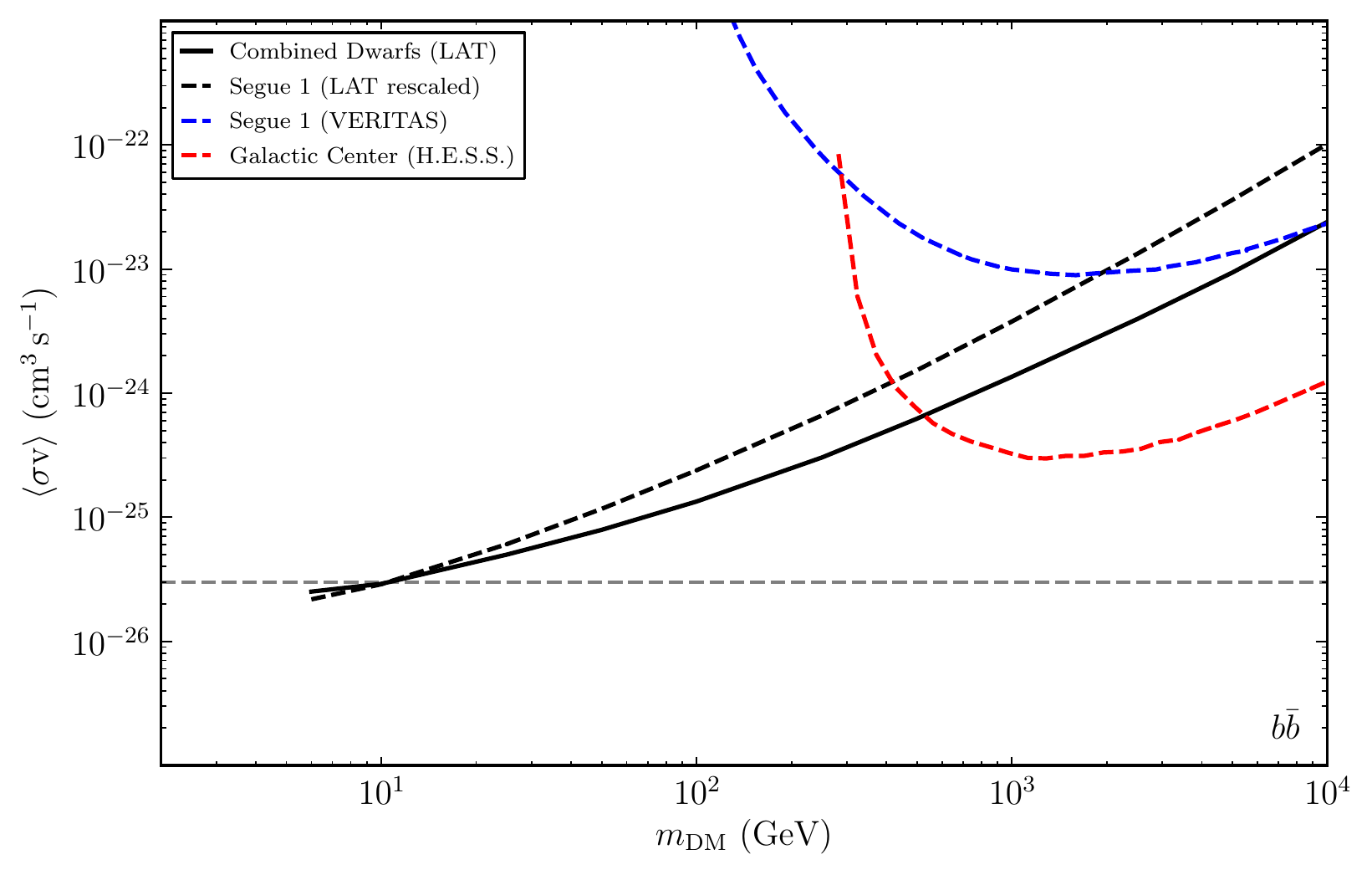}
  \caption{{\capEight}}\label{fig:fig8}
\end{figure}

\section*{}
\label{sec:tables}

\begin{turnpage}
\begin{table}[h] \scriptsize
\caption{ \label{tab:dsphs} Properties of Milky Way dwarf spheroidal satellite galaxies.}

\begin{ruledtabular}
\begin{tabular}{ l c c c c c c c c}
Name                      & GLON & GLAT & Distance & $\overline{\log_{10}({\rm J^{NFW}})}$\footnotemark[1] & $\overline{\log_{10}(\alphas^{\rm NFW})}$ & $\overline{\log_{10}({\rm J^{Burkert}})}$\footnotemark[1] & $\overline{\log_{10}(\alphas^{\rm Burkert})}$ & Reference \\
                          & ($\deg$) & ($\deg$) & (\kpc) & ($\log_{10}[\GeV^2 \cm^{-5} \sr]$) & ($\log_{10}[\deg]$) & ($\log_{10}[\GeV^2 \cm^{-5} \sr]$) & ($\log_{10}[\deg]$) &  \\
\hline
Bootes I                  & 358.1  & 69.6   & 66     & $18.8 \pm 0.22$ & $-0.6 \pm 0.3$ & $18.6 \pm 0.17$ & $-1.1 \pm 0.2$ & ~\cite{Dall'Ora:2006pt} \\
Bootes II                 & 353.7  & 68.9   & 42     &              -- &     -- &              -- &     -- &     -- \\
Bootes III                & 35.4   & 75.4   & 47     &              -- &     -- &              -- &     -- &     -- \\
Canes Venatici I          & 74.3   & 79.8   & 218    & $17.7 \pm 0.26$ & $-1.3 \pm 0.2$ & $17.6 \pm 0.17$ & $-1.6 \pm 0.1$ & ~\cite{Simon:2007dq} \\
Canes Venatici II         & 113.6  & 82.7   & 160    & $17.9 \pm 0.25$ & $-1.1 \pm 0.4$ & $17.8 \pm 0.19$ & $-1.6 \pm 0.2$ & ~\cite{Simon:2007dq} \\
Canis Major               & 240.0  & -8.0   & 7      &              -- &     -- &              -- &     -- &     -- \\
Carina                    & 260.1  & -22.2  & 105    & $18.1 \pm 0.23$ & $-1.0 \pm 0.3$ & $18.1 \pm 0.16$ & $-1.5 \pm 0.1$ & ~\cite{Walker:2008ax} \\
Coma Berenices            & 241.9  & 83.6   & 44     & $19.0 \pm 0.25$ & $-0.6 \pm 0.5$ & $18.9 \pm 0.21$ & $-1.1 \pm 0.3$ & ~\cite{Simon:2007dq} \\
Draco                     & 86.4   & 34.7   & 76     & $18.8 \pm 0.16$ & $-0.6 \pm 0.2$ & $18.7 \pm 0.17$ & $-1.0 \pm 0.2$ & ~\cite{Munoz:2005be} \\
Fornax                    & 237.1  & -65.7  & 147    & $18.2 \pm 0.21$ & $-0.8 \pm 0.2$ & $18.1 \pm 0.22$ & $-1.1 \pm 0.1$ & ~\cite{Walker:2008ax} \\
Hercules                  & 28.7   & 36.9   & 132    & $18.1 \pm 0.25$ & $-1.1 \pm 0.4$ & $17.9 \pm 0.19$ & $-1.6 \pm 0.2$ & ~\cite{Simon:2007dq} \\
Leo I                     & 226.0  & 49.1   & 254    & $17.7 \pm 0.18$ & $-1.1 \pm 0.3$ & $17.6 \pm 0.17$ & $-1.5 \pm 0.2$ & ~\cite{Mateo:2007xh} \\
Leo II                    & 220.2  & 67.2   & 233    & $17.6 \pm 0.18$ & $-1.1 \pm 0.5$ & $17.5 \pm 0.15$ & $-1.9 \pm 0.5$ & ~\cite{Koch:2007ye} \\
Leo IV                    & 265.4  & 56.5   & 154    & $17.9 \pm 0.28$ & $-1.1 \pm 0.4$ & $17.8 \pm 0.21$ & $-1.6 \pm 0.2$ & ~\cite{Simon:2007dq} \\
Leo V                     & 261.9  & 58.5   & 178    &              -- &     -- &              -- &     -- &     -- \\
Pisces II                 & 79.2   & -47.1  & 182    &              -- &     -- &              -- &     -- &     -- \\
Sagittarius               & 5.6    & -14.2  & 26     &              -- &     -- &              -- &     -- &     -- \\
Sculptor                  & 287.5  & -83.2  & 86     & $18.6 \pm 0.18$ & $-0.6 \pm 0.3$ & $18.5 \pm 0.17$ & $-1.1 \pm 0.2$ & ~\cite{Walker:2008ax} \\
Segue 1                   & 220.5  & 50.4   & 23     & $19.5 \pm 0.29$ & $-0.4 \pm 0.5$ & $19.4 \pm 0.24$ & $-0.9 \pm 0.2$ & ~\cite{Simon:2010ek} \\
Segue 2                   & 149.4  & -38.1  & 35     &              -- &     -- &              -- &     -- &     -- \\
Sextans                   & 243.5  & 42.3   & 86     & $18.4 \pm 0.27$ & $-0.9 \pm 0.2$ & $18.4 \pm 0.16$ & $-1.2 \pm 0.1$ & ~\cite{Walker:2008ax} \\
Ursa Major I              & 159.4  & 54.4   & 97     & $18.3 \pm 0.24$ & $-1.0 \pm 0.3$ & $18.2 \pm 0.18$ & $-1.4 \pm 0.2$ & ~\cite{Simon:2007dq} \\
Ursa Major II             & 152.5  & 37.4   & 32     & $19.3 \pm 0.28$ & $-0.5 \pm 0.4$ & $19.2 \pm 0.21$ & $-0.9 \pm 0.2$ & ~\cite{Simon:2007dq} \\
Ursa Minor                & 105.0  & 44.8   & 76     & $18.8 \pm 0.19$ & $-0.5 \pm 0.2$ & $18.7 \pm 0.20$ & $-0.9 \pm 0.2$ & ~\cite{Munoz:2005be} \\
Willman 1                 & 158.6  & 56.8   & 38     & $19.1 \pm 0.31$ & $-0.6 \pm 0.5$ & $19.0 \pm 0.28$ & $-1.1 \pm 0.2$ & ~\cite{Willman:2010gy} \\

\footnotetext[1]{ \Jfactors are calculated over a solid angle of $\Delta\Omega \sim 2.4 \times 10^{-4} \sr$.}
\end{tabular}
\end{ruledtabular}

\end{table}
\end{turnpage}



\begin{turnpage}

\begin{table}[h] \scriptsize
\caption{\label{tab:ee_limits} 95\% CL limits on the annihilation cross section for the \ee channel ($\mathrm{cm}^3\,\second^{-1}$). }

\begin{ruledtabular}
\begin{tabular}{ l  c c c c c c c c c c c c}
 & 2\GeV  & 5\GeV  & 10\GeV  & 25\GeV  & 50\GeV  & 100\GeV  & 250\GeV  & 500\GeV  & 1000\GeV  & 2500\GeV  & 5000\GeV  & 10000\GeV \\
\hline
Bootes I                   & 1.49e-25  & 2.12e-25  & 3.26e-25  & 1.68e-24  & 3.80e-24  & 9.81e-24  & 3.74e-23  & 1.09e-22  & 3.71e-22  & 2.01e-21  & 7.45e-21  & 2.82e-20  \\

Bootes II                  & --  & --  & --  & --  & --  & --  & --  & --  & --  & --  & --  & --  \\

Bootes III                 & --  & --  & --  & --  & --  & --  & --  & --  & --  & --  & --  & --  \\

Canes Venatici I           & 1.35e-24  & 1.20e-24  & 5.51e-24  & 1.51e-23  & 3.58e-23  & 9.33e-23  & 4.18e-22  & 1.28e-21  & 4.46e-21  & 2.47e-20  & 9.23e-20  & 3.50e-19  \\

Canes Venatici II          & 2.42e-24  & 2.21e-24  & 3.16e-24  & 5.42e-24  & 1.20e-23  & 3.05e-23  & 1.23e-22  & 3.84e-22  & 1.34e-21  & 7.47e-21  & 2.79e-20  & 1.06e-19  \\

Canis Major                & --  & --  & --  & --  & --  & --  & --  & --  & --  & --  & --  & --  \\

Carina                     & 1.06e-24  & 9.76e-25  & 9.94e-25  & 2.10e-24  & 4.99e-24  & 3.65e-23  & 1.54e-22  & 4.86e-22  & 1.71e-21  & 9.53e-21  & 3.57e-20  & 1.35e-19  \\

Coma Berenices             & 7.90e-26  & 8.25e-26  & 1.43e-25  & 3.72e-25  & 8.71e-25  & 2.34e-24  & 9.75e-24  & 3.08e-23  & 1.09e-22  & 6.05e-22  & 2.26e-21  & 8.60e-21  \\

Draco                      & 6.26e-26  & 1.41e-25  & 2.11e-25  & 5.23e-25  & 1.35e-24  & 3.21e-24  & 1.18e-23  & 3.46e-23  & 1.19e-22  & 6.49e-22  & 2.41e-21  & 9.14e-21  \\

Fornax                     & 3.76e-25  & 1.05e-24  & 1.43e-24  & 3.80e-24  & 1.01e-23  & 2.34e-23  & 8.50e-23  & 2.49e-22  & 8.53e-22  & 4.66e-21  & 1.73e-20  & 6.56e-20  \\

Hercules                   & 2.33e-24  & 3.34e-24  & 7.95e-24  & 1.71e-23  & 3.40e-23  & 7.75e-23  & 2.68e-22  & 7.54e-22  & 2.51e-21  & 1.34e-20  & 4.96e-20  & 1.87e-19  \\

Leo I                      & 1.75e-24  & 3.14e-24  & 4.01e-24  & 9.14e-24  & 1.90e-23  & 5.03e-23  & 2.19e-22  & 6.79e-22  & 2.38e-21  & 1.32e-20  & 4.93e-20  & 1.87e-19  \\

Leo II                     & 2.78e-24  & 1.80e-24  & 2.26e-24  & 5.52e-24  & 1.48e-23  & 4.12e-23  & 1.79e-22  & 5.74e-22  & 2.03e-21  & 1.14e-20  & 4.26e-20  & 1.62e-19  \\

Leo IV                     & 5.89e-25  & 7.87e-25  & 1.22e-24  & 3.42e-24  & 8.82e-24  & 2.54e-23  & 1.13e-22  & 3.68e-22  & 1.31e-21  & 7.37e-21  & 2.77e-20  & 1.05e-19  \\

Leo V                      & --  & --  & --  & --  & --  & --  & --  & --  & --  & --  & --  & --  \\

Pisces II                  & --  & --  & --  & --  & --  & --  & --  & --  & --  & --  & --  & --  \\

Sagittarius                & --  & --  & --  & --  & --  & --  & --  & --  & --  & --  & --  & --  \\

Sculptor                   & 1.36e-25  & 3.24e-25  & 1.16e-24  & 2.76e-24  & 5.98e-24  & 1.36e-23  & 4.64e-23  & 1.27e-22  & 4.23e-22  & 2.27e-21  & 8.39e-21  & 3.17e-20  \\

Segue 1                    & 7.38e-26  & 9.73e-26  & 2.81e-25  & 7.46e-25  & 1.57e-24  & 3.85e-24  & 1.30e-23  & 3.46e-23  & 1.13e-22  & 5.97e-22  & 2.19e-21  & 8.25e-21  \\

Segue 2                    & --  & --  & --  & --  & --  & --  & --  & --  & --  & --  & --  & --  \\

Sextans                    & 1.18e-24  & 7.97e-25  & 9.34e-25  & 1.76e-24  & 3.94e-24  & 1.04e-23  & 4.32e-23  & 1.39e-22  & 4.91e-22  & 2.75e-21  & 1.03e-20  & 3.92e-20  \\

Ursa Major I               & 1.35e-25  & 2.96e-25  & 5.08e-25  & 1.24e-24  & 3.02e-24  & 9.20e-24  & 3.79e-23  & 1.18e-22  & 4.13e-22  & 2.29e-21  & 8.57e-21  & 3.25e-20  \\

Ursa Major II              & 9.92e-26  & 1.49e-25  & 2.73e-25  & 3.92e-25  & 6.69e-25  & 2.42e-24  & 8.47e-24  & 2.41e-23  & 8.20e-23  & 4.45e-22  & 1.65e-21  & 6.24e-21  \\

Ursa Minor                 & 5.80e-26  & 8.27e-26  & 1.30e-25  & 4.03e-25  & 9.83e-25  & 2.50e-24  & 9.78e-24  & 2.98e-23  & 1.03e-22  & 5.71e-22  & 2.13e-21  & 8.08e-21  \\

Willman 1                  & 2.00e-25  & 2.79e-25  & 4.52e-25  & 1.10e-24  & 2.65e-24  & 6.45e-24  & 2.36e-23  & 6.81e-23  & 2.29e-22  & 1.23e-21  & 4.56e-21  & 1.72e-20  \\

Combined                   & 4.24e-26  & 5.07e-26  & 1.03e-25  & 2.16e-25  & 3.71e-25  & 8.71e-25  & 2.26e-24  & 5.68e-24  & 1.85e-23  & 9.78e-23  & 3.59e-22  & 1.35e-21  \\

\end{tabular}
\end{ruledtabular}
    
\end{table}

\begin{table}[h] \scriptsize
\caption{\label{tab:mumu_limits} 95\% CL limits on the annihilation cross section for the \mumu channel ($\mathrm{cm}^3\,\second^{-1}$). }

\begin{ruledtabular}
\begin{tabular}{ l  c c c c c c c c c c c c}
 & 2\GeV  & 5\GeV  & 10\GeV  & 25\GeV  & 50\GeV  & 100\GeV  & 250\GeV  & 500\GeV  & 1000\GeV  & 2500\GeV  & 5000\GeV  & 10000\GeV \\
\hline
Bootes I                   & 3.59e-25  & 5.05e-25  & 6.64e-25  & 2.64e-24  & 5.69e-24  & 1.39e-23  & 5.04e-23  & 1.43e-22  & 4.59e-22  & 2.36e-21  & 8.49e-21  & 3.14e-20  \\

Bootes II                  & --  & --  & --  & --  & --  & --  & --  & --  & --  & --  & --  & --  \\

Bootes III                 & --  & --  & --  & --  & --  & --  & --  & --  & --  & --  & --  & --  \\

Canes Venatici I           & 3.66e-24  & 2.74e-24  & 9.59e-24  & 2.38e-23  & 5.35e-23  & 1.32e-22  & 5.44e-22  & 1.62e-21  & 5.40e-21  & 2.86e-20  & 1.04e-19  & 3.89e-19  \\

Canes Venatici II          & 6.31e-24  & 5.51e-24  & 6.77e-24  & 9.63e-24  & 1.87e-23  & 4.39e-23  & 1.64e-22  & 4.87e-22  & 1.62e-21  & 8.62e-21  & 3.15e-20  & 1.18e-19  \\

Canis Major                & --  & --  & --  & --  & --  & --  & --  & --  & --  & --  & --  & --  \\

Carina                     & 2.81e-24  & 2.53e-24  & 2.10e-24  & 3.57e-24  & 7.57e-24  & 4.99e-23  & 1.99e-22  & 6.09e-22  & 2.05e-21  & 1.10e-20  & 4.02e-20  & 1.50e-19  \\

Coma Berenices             & 2.02e-25  & 1.77e-25  & 2.61e-25  & 6.01e-25  & 1.31e-24  & 3.28e-24  & 1.28e-23  & 3.88e-23  & 1.30e-22  & 6.96e-22  & 2.55e-21  & 9.54e-21  \\

Draco                      & 1.87e-25  & 2.83e-25  & 3.88e-25  & 8.60e-25  & 2.00e-24  & 4.61e-24  & 1.60e-23  & 4.49e-23  & 1.46e-22  & 7.57e-22  & 2.74e-21  & 1.02e-20  \\

Fornax                     & 9.70e-25  & 2.21e-24  & 2.81e-24  & 6.30e-24  & 1.49e-23  & 3.37e-23  & 1.15e-22  & 3.23e-22  & 1.05e-21  & 5.43e-21  & 1.97e-20  & 7.31e-20  \\

Hercules                   & 6.53e-24  & 7.46e-24  & 1.49e-23  & 2.93e-23  & 5.47e-23  & 1.17e-22  & 3.77e-22  & 1.02e-21  & 3.17e-21  & 1.59e-20  & 5.69e-20  & 2.09e-19  \\

Leo I                      & 5.62e-24  & 7.04e-24  & 8.27e-24  & 1.54e-23  & 2.96e-23  & 7.14e-23  & 2.88e-22  & 8.58e-22  & 2.87e-21  & 1.52e-20  & 5.56e-20  & 2.08e-19  \\

Leo II                     & 8.61e-24  & 4.91e-24  & 4.55e-24  & 9.03e-24  & 2.15e-23  & 5.64e-23  & 2.31e-22  & 7.14e-22  & 2.43e-21  & 1.30e-20  & 4.79e-20  & 1.79e-19  \\

Leo IV                     & 1.87e-24  & 1.65e-24  & 2.24e-24  & 5.38e-24  & 1.28e-23  & 3.46e-23  & 1.44e-22  & 4.55e-22  & 1.56e-21  & 8.43e-21  & 3.10e-20  & 1.16e-19  \\

Leo V                      & --  & --  & --  & --  & --  & --  & --  & --  & --  & --  & --  & --  \\

Pisces II                  & --  & --  & --  & --  & --  & --  & --  & --  & --  & --  & --  & --  \\

Sagittarius                & --  & --  & --  & --  & --  & --  & --  & --  & --  & --  & --  & --  \\

Sculptor                   & 3.99e-25  & 7.06e-25  & 2.07e-24  & 4.55e-24  & 9.38e-24  & 2.06e-23  & 6.63e-23  & 1.72e-22  & 5.34e-22  & 2.69e-21  & 9.61e-21  & 3.55e-20  \\

Segue 1                    & 1.84e-25  & 2.27e-25  & 5.06e-25  & 1.24e-24  & 2.48e-24  & 5.71e-24  & 1.85e-23  & 4.79e-23  & 1.45e-22  & 7.13e-22  & 2.53e-21  & 9.28e-21  \\

Segue 2                    & --  & --  & --  & --  & --  & --  & --  & --  & --  & --  & --  & --  \\

Sextans                    & 2.97e-24  & 2.12e-24  & 2.02e-24  & 3.03e-24  & 6.08e-24  & 1.47e-23  & 5.67e-23  & 1.74e-22  & 5.88e-22  & 3.16e-21  & 1.16e-20  & 4.34e-20  \\

Ursa Major I               & 3.80e-25  & 6.19e-25  & 9.30e-25  & 2.01e-24  & 4.50e-24  & 1.26e-23  & 4.94e-23  & 1.48e-22  & 4.97e-22  & 2.64e-21  & 9.66e-21  & 3.61e-20  \\

Ursa Major II              & 2.64e-25  & 3.55e-25  & 5.52e-25  & 7.49e-25  & 1.13e-24  & 3.50e-24  & 1.16e-23  & 3.17e-23  & 1.01e-22  & 5.21e-22  & 1.88e-21  & 6.96e-21  \\

Ursa Minor                 & 1.61e-25  & 1.65e-25  & 2.34e-25  & 6.22e-25  & 1.45e-24  & 3.53e-24  & 1.30e-23  & 3.80e-23  & 1.25e-22  & 6.62e-22  & 2.41e-21  & 8.98e-21  \\

Willman 1                  & 5.43e-25  & 6.50e-25  & 9.06e-25  & 1.85e-24  & 4.07e-24  & 9.44e-24  & 3.26e-23  & 9.08e-23  & 2.87e-22  & 1.46e-21  & 5.21e-21  & 1.93e-20  \\

Combined                   & 1.16e-25  & 1.22e-25  & 2.01e-25  & 3.77e-25  & 6.33e-25  & 1.38e-24  & 3.38e-24  & 7.92e-24  & 2.37e-23  & 1.17e-22  & 4.13e-22  & 1.52e-21  \\

\end{tabular}
\end{ruledtabular}
    
\end{table}

\begin{table}[h] \scriptsize
\caption{\label{tab:tautau_limits} 95\% CL limits on the annihilation cross section for the \tautau channel ($\mathrm{cm}^3\,\second^{-1}$). }

\begin{ruledtabular}
\begin{tabular}{ l  c c c c c c c c c c c c}
 & 2\GeV  & 5\GeV  & 10\GeV  & 25\GeV  & 50\GeV  & 100\GeV  & 250\GeV  & 500\GeV  & 1000\GeV  & 2500\GeV  & 5000\GeV  & 10000\GeV \\
\hline
Bootes I                   & 4.04e-26  & 6.74e-26  & 9.99e-26  & 2.17e-25  & 6.34e-25  & 1.78e-24  & 7.46e-24  & 2.19e-23  & 6.71e-23  & 4.02e-22  & 2.01e-21  & 1.15e-20  \\

Bootes II                  & --  & --  & --  & --  & --  & --  & --  & --  & --  & --  & --  & --  \\

Bootes III                 & --  & --  & --  & --  & --  & --  & --  & --  & --  & --  & --  & --  \\

Canes Venatici I           & 5.54e-25  & 4.20e-25  & 7.48e-25  & 2.62e-24  & 6.66e-24  & 1.78e-23  & 6.86e-23  & 2.22e-22  & 8.41e-22  & 6.05e-21  & 3.17e-20  & 1.81e-19  \\

Canes Venatici II          & 5.25e-25  & 8.30e-25  & 9.41e-25  & 1.15e-24  & 2.00e-24  & 4.86e-24  & 2.16e-23  & 7.62e-23  & 2.97e-22  & 2.17e-21  & 1.12e-20  & 6.21e-20  \\

Canis Major                & --  & --  & --  & --  & --  & --  & --  & --  & --  & --  & --  & --  \\

Carina                     & 2.87e-25  & 4.24e-25  & 3.54e-25  & 4.10e-25  & 8.23e-25  & 3.96e-24  & 2.90e-23  & 1.06e-22  & 4.15e-22  & 2.98e-21  & 1.52e-20  & 8.25e-20  \\

Coma Berenices             & 3.15e-26  & 2.26e-26  & 2.99e-26  & 7.01e-26  & 1.61e-25  & 4.24e-25  & 1.88e-24  & 6.58e-24  & 2.55e-23  & 1.85e-22  & 9.56e-22  & 5.25e-21  \\

Draco                      & 3.35e-26  & 1.95e-26  & 4.18e-26  & 1.04e-25  & 2.26e-25  & 5.78e-25  & 2.23e-24  & 6.80e-24  & 2.37e-23  & 1.59e-22  & 8.12e-22  & 4.53e-21  \\

Fornax                     & 1.94e-25  & 2.37e-25  & 3.51e-25  & 7.10e-25  & 1.47e-24  & 4.01e-24  & 1.65e-23  & 5.05e-23  & 1.74e-22  & 1.16e-21  & 5.86e-21  & 3.26e-20  \\

Hercules                   & 7.94e-25  & 8.17e-25  & 1.39e-24  & 3.52e-24  & 6.82e-24  & 1.42e-23  & 4.23e-23  & 1.03e-22  & 3.20e-22  & 2.09e-21  & 1.10e-20  & 6.56e-20  \\

Leo I                      & 9.35e-25  & 8.84e-25  & 1.10e-24  & 1.70e-24  & 3.34e-24  & 8.25e-24  & 3.83e-23  & 1.41e-22  & 5.51e-22  & 3.90e-21  & 1.98e-20  & 1.08e-19  \\

Leo II                     & 1.51e-24  & 9.68e-25  & 6.13e-25  & 9.51e-25  & 2.22e-24  & 6.75e-24  & 3.55e-23  & 1.32e-22  & 5.18e-22  & 3.74e-21  & 1.91e-20  & 1.04e-19  \\

Leo IV                     & 5.03e-25  & 2.06e-25  & 2.54e-25  & 5.96e-25  & 1.49e-24  & 4.47e-24  & 2.28e-23  & 8.53e-23  & 3.44e-22  & 2.56e-21  & 1.32e-20  & 7.15e-20  \\

Leo V                      & --  & --  & --  & --  & --  & --  & --  & --  & --  & --  & --  & --  \\

Pisces II                  & --  & --  & --  & --  & --  & --  & --  & --  & --  & --  & --  & --  \\

Sagittarius                & --  & --  & --  & --  & --  & --  & --  & --  & --  & --  & --  & --  \\

Sculptor                   & 4.51e-26  & 8.13e-26  & 2.03e-25  & 5.42e-25  & 1.17e-24  & 2.48e-24  & 6.33e-24  & 1.70e-23  & 5.84e-23  & 4.05e-22  & 2.13e-21  & 1.24e-20  \\

Segue 1                    & 1.19e-26  & 2.97e-26  & 4.91e-26  & 1.38e-25  & 2.96e-25  & 6.70e-25  & 1.89e-24  & 4.40e-24  & 1.34e-23  & 8.73e-23  & 4.56e-22  & 2.69e-21  \\

Segue 2                    & --  & --  & --  & --  & --  & --  & --  & --  & --  & --  & --  & --  \\

Sextans                    & 1.89e-25  & 3.51e-25  & 3.35e-25  & 3.58e-25  & 6.77e-25  & 1.72e-24  & 7.83e-24  & 2.82e-23  & 1.13e-22  & 8.61e-22  & 4.55e-21  & 2.53e-20  \\

Ursa Major I               & 7.77e-26  & 7.12e-26  & 1.06e-25  & 2.40e-25  & 5.40e-25  & 1.54e-24  & 7.63e-24  & 2.69e-23  & 1.00e-22  & 6.90e-22  & 3.48e-21  & 1.90e-20  \\

Ursa Major II              & 3.30e-26  & 4.68e-26  & 7.17e-26  & 1.05e-25  & 1.43e-25  & 3.25e-25  & 1.44e-24  & 4.72e-24  & 1.63e-23  & 1.05e-22  & 5.17e-22  & 2.86e-21  \\

Ursa Minor                 & 2.23e-26  & 1.66e-26  & 2.49e-26  & 6.60e-26  & 1.76e-25  & 4.91e-25  & 2.01e-24  & 6.36e-24  & 2.29e-23  & 1.57e-22  & 8.04e-22  & 4.45e-21  \\

Willman 1                  & 7.05e-26  & 7.91e-26  & 1.09e-25  & 1.93e-25  & 4.15e-25  & 1.07e-24  & 3.65e-24  & 9.61e-24  & 3.13e-23  & 2.11e-22  & 1.11e-21  & 6.53e-21  \\

Combined                   & 1.37e-26  & 1.65e-26  & 2.37e-26  & 4.46e-26  & 8.08e-26  & 1.50e-25  & 3.55e-25  & 9.00e-25  & 2.81e-24  & 1.78e-23  & 8.93e-23  & 5.06e-22  \\

\end{tabular}
\end{ruledtabular}
    
\end{table}

\begin{table}[h] \scriptsize
\caption{\label{tab:uu_limits} 95\% CL limits on the annihilation cross section for the \uubar channel ($\mathrm{cm}^3\,\second^{-1}$). }

\begin{ruledtabular}
\begin{tabular}{ l  c c c c c c c c c c c c}
 & 2\GeV  & 5\GeV  & 10\GeV  & 25\GeV  & 50\GeV  & 100\GeV  & 250\GeV  & 500\GeV  & 1000\GeV  & 2500\GeV  & 5000\GeV  & 10000\GeV \\
\hline
Bootes I                   & 2.39e-26  & 4.39e-26  & 7.98e-26  & 1.54e-25  & 3.03e-25  & 7.22e-25  & 2.22e-24  & 5.35e-24  & 1.34e-23  & 4.82e-23  & 1.32e-22  & 3.74e-22  \\

Bootes II                  & --  & --  & --  & --  & --  & --  & --  & --  & --  & --  & --  & --  \\

Bootes III                 & --  & --  & --  & --  & --  & --  & --  & --  & --  & --  & --  & --  \\

Canes Venatici I           & 3.22e-25  & 4.31e-25  & 5.41e-25  & 1.42e-24  & 3.06e-24  & 6.71e-24  & 2.03e-23  & 5.01e-23  & 1.30e-22  & 4.93e-22  & 1.38e-21  & 4.01e-21  \\

Canes Venatici II          & 3.29e-25  & 6.94e-25  & 1.03e-24  & 1.68e-24  & 2.48e-24  & 3.89e-24  & 8.94e-24  & 1.96e-23  & 4.65e-23  & 1.61e-22  & 4.34e-22  & 1.22e-21  \\

Canis Major                & --  & --  & --  & --  & --  & --  & --  & --  & --  & --  & --  & --  \\

Carina                     & 1.68e-25  & 3.46e-25  & 5.06e-25  & 6.42e-25  & 7.98e-25  & 1.32e-24  & 4.87e-24  & 1.64e-23  & 4.58e-23  & 1.74e-22  & 4.90e-22  & 1.42e-21  \\

Coma Berenices             & 1.72e-26  & 2.24e-26  & 2.89e-26  & 4.96e-26  & 9.00e-26  & 1.81e-25  & 5.17e-25  & 1.25e-24  & 3.17e-24  & 1.17e-23  & 3.26e-23  & 9.41e-23  \\

Draco                      & 1.93e-26  & 2.18e-26  & 3.21e-26  & 6.88e-26  & 1.30e-25  & 2.65e-25  & 7.66e-25  & 1.83e-24  & 4.52e-24  & 1.59e-23  & 4.29e-23  & 1.20e-22  \\

Fornax                     & 1.05e-25  & 1.55e-25  & 2.88e-25  & 5.69e-25  & 1.04e-24  & 2.06e-24  & 5.86e-24  & 1.38e-23  & 3.34e-23  & 1.15e-22  & 3.09e-22  & 8.60e-22  \\

Hercules                   & 4.65e-25  & 7.57e-25  & 1.17e-24  & 2.76e-24  & 5.14e-24  & 9.77e-24  & 2.50e-23  & 5.52e-23  & 1.29e-22  & 4.31e-22  & 1.12e-21  & 3.05e-21  \\

Leo I                      & 5.56e-25  & 8.11e-25  & 1.16e-24  & 1.91e-24  & 2.99e-24  & 5.24e-24  & 1.33e-23  & 3.05e-23  & 7.53e-23  & 2.70e-22  & 7.38e-22  & 2.10e-21  \\

Leo II                     & 8.50e-25  & 1.11e-24  & 1.29e-24  & 1.29e-24  & 1.70e-24  & 3.02e-24  & 8.45e-24  & 2.06e-23  & 5.31e-23  & 1.99e-22  & 5.62e-22  & 1.64e-21  \\

Leo IV                     & 2.72e-25  & 2.48e-25  & 2.70e-25  & 4.24e-25  & 7.57e-25  & 1.57e-24  & 4.78e-24  & 1.21e-23  & 3.19e-23  & 1.23e-22  & 3.51e-22  & 1.04e-21  \\

Leo V                      & --  & --  & --  & --  & --  & --  & --  & --  & --  & --  & --  & --  \\

Pisces II                  & --  & --  & --  & --  & --  & --  & --  & --  & --  & --  & --  & --  \\

Sagittarius                & --  & --  & --  & --  & --  & --  & --  & --  & --  & --  & --  & --  \\

Sculptor                   & 2.73e-26  & 5.30e-26  & 1.06e-25  & 3.19e-25  & 6.57e-25  & 1.38e-24  & 3.93e-24  & 9.20e-24  & 2.24e-23  & 7.72e-23  & 2.03e-22  & 5.48e-22  \\

Segue 1                    & 7.53e-27  & 2.04e-26  & 3.76e-26  & 8.97e-26  & 1.84e-25  & 3.83e-25  & 1.07e-24  & 2.49e-24  & 6.04e-24  & 2.07e-23  & 5.44e-23  & 1.48e-22  \\

Segue 2                    & --  & --  & --  & --  & --  & --  & --  & --  & --  & --  & --  & --  \\

Sextans                    & 1.24e-25  & 2.94e-25  & 4.33e-25  & 6.14e-25  & 7.52e-25  & 1.15e-24  & 2.75e-24  & 6.20e-24  & 1.51e-23  & 5.39e-23  & 1.48e-22  & 4.23e-22  \\

Ursa Major I               & 4.27e-26  & 5.59e-26  & 8.67e-26  & 1.66e-25  & 3.03e-25  & 6.13e-25  & 1.81e-24  & 4.51e-24  & 1.17e-23  & 4.38e-23  & 1.22e-22  & 3.55e-22  \\

Ursa Major II              & 1.80e-26  & 3.44e-26  & 6.06e-26  & 1.19e-25  & 1.89e-25  & 3.12e-25  & 7.39e-25  & 1.62e-24  & 3.76e-24  & 1.23e-23  & 3.19e-23  & 8.69e-23  \\

Ursa Minor                 & 1.27e-26  & 1.73e-26  & 2.26e-26  & 4.16e-26  & 7.98e-26  & 1.73e-25  & 5.32e-25  & 1.30e-24  & 3.31e-24  & 1.21e-23  & 3.34e-23  & 9.57e-23  \\

Willman 1                  & 4.06e-26  & 6.59e-26  & 1.06e-25  & 1.99e-25  & 3.32e-25  & 6.31e-25  & 1.73e-24  & 4.02e-24  & 9.77e-24  & 3.40e-23  & 9.11e-23  & 2.53e-22  \\

Combined                   & 7.92e-27  & 1.40e-26  & 2.23e-26  & 4.24e-26  & 7.18e-26  & 1.31e-25  & 3.25e-25  & 7.01e-25  & 1.57e-24  & 4.75e-24  & 1.13e-23  & 2.83e-23  \\

\end{tabular}
\end{ruledtabular}
    
\end{table}

\begin{table}[h] \scriptsize
\caption{\label{tab:bb_limits} 95\% CL limits on the annihilation cross section for the \bbbar channel ($\mathrm{cm}^3\,\second^{-1}$). }

\begin{ruledtabular}
\begin{tabular}{ l  c c c c c c c c c c c}
 & 6\GeV  & 10\GeV  & 25\GeV  & 50\GeV  & 100\GeV  & 250\GeV  & 500\GeV  & 1000\GeV  & 2500\GeV  & 5000\GeV  & 10000\GeV \\
\hline
Bootes I                   & 7.83e-26  & 9.41e-26  & 1.78e-25  & 3.08e-25  & 6.29e-25  & 1.94e-24  & 4.62e-24  & 1.16e-23  & 4.30e-23  & 1.18e-22  & 3.40e-22  \\

Bootes II                  & --  & --  & --  & --  & --  & --  & --  & --  & --  & --  & --  \\

Bootes III                 & --  & --  & --  & --  & --  & --  & --  & --  & --  & --  & --  \\

Canes Venatici I           & 8.66e-25  & 7.99e-25  & 1.40e-24  & 2.88e-24  & 6.10e-24  & 1.80e-23  & 4.38e-23  & 1.14e-22  & 4.31e-22  & 1.24e-21  & 3.69e-21  \\

Canes Venatici II          & 1.18e-24  & 1.38e-24  & 2.13e-24  & 3.00e-24  & 4.42e-24  & 8.26e-24  & 1.67e-23  & 3.90e-23  & 1.39e-22  & 3.86e-22  & 1.12e-21  \\

Canis Major                & --  & --  & --  & --  & --  & --  & --  & --  & --  & --  & --  \\

Carina                     & 5.83e-25  & 6.87e-25  & 9.33e-25  & 1.08e-24  & 1.42e-24  & 3.53e-24  & 1.11e-23  & 3.72e-23  & 1.54e-22  & 4.47e-22  & 1.33e-21  \\

Coma Berenices             & 4.59e-26  & 4.21e-26  & 5.86e-26  & 9.38e-26  & 1.73e-25  & 4.65e-25  & 1.09e-24  & 2.76e-24  & 1.03e-23  & 2.94e-23  & 8.67e-23  \\

Draco                      & 4.87e-26  & 4.52e-26  & 7.47e-26  & 1.32e-25  & 2.53e-25  & 6.82e-25  & 1.62e-24  & 3.95e-24  & 1.40e-23  & 3.83e-23  & 1.09e-22  \\

Fornax                     & 2.97e-25  & 3.49e-25  & 6.35e-25  & 1.09e-24  & 2.00e-24  & 5.28e-24  & 1.18e-23  & 2.89e-23  & 1.02e-22  & 2.76e-22  & 7.78e-22  \\

Hercules                   & 1.42e-24  & 1.55e-24  & 2.95e-24  & 5.33e-24  & 9.74e-24  & 2.32e-23  & 4.94e-23  & 1.12e-22  & 3.66e-22  & 9.68e-22  & 2.69e-21  \\

Leo I                      & 1.57e-24  & 1.63e-24  & 2.37e-24  & 3.48e-24  & 5.48e-24  & 1.22e-23  & 2.62e-23  & 6.32e-23  & 2.34e-22  & 6.61e-22  & 1.94e-21  \\

Leo II                     & 2.29e-24  & 2.09e-24  & 2.16e-24  & 2.16e-24  & 3.06e-24  & 7.30e-24  & 1.72e-23  & 4.53e-23  & 1.77e-22  & 5.14e-22  & 1.53e-21  \\

Leo IV                     & 5.70e-25  & 4.39e-25  & 5.16e-25  & 7.93e-25  & 1.48e-24  & 4.17e-24  & 1.04e-23  & 2.76e-23  & 1.10e-22  & 3.21e-22  & 9.68e-22  \\

Leo V                      & --  & --  & --  & --  & --  & --  & --  & --  & --  & --  & --  \\

Pisces II                  & --  & --  & --  & --  & --  & --  & --  & --  & --  & --  & --  \\

Sagittarius                & --  & --  & --  & --  & --  & --  & --  & --  & --  & --  & --  \\

Sculptor                   & 8.90e-26  & 1.18e-25  & 3.05e-25  & 6.34e-25  & 1.31e-24  & 3.59e-24  & 8.18e-24  & 1.94e-23  & 6.55e-23  & 1.74e-22  & 4.81e-22  \\

Segue 1                    & 3.13e-26  & 4.36e-26  & 9.39e-26  & 1.80e-25  & 3.62e-25  & 9.68e-25  & 2.17e-24  & 5.22e-24  & 1.77e-23  & 4.71e-23  & 1.31e-22  \\

Segue 2                    & --  & --  & --  & --  & --  & --  & --  & --  & --  & --  & --  \\

Sextans                    & 4.82e-25  & 5.76e-25  & 8.54e-25  & 1.04e-24  & 1.30e-24  & 2.52e-24  & 5.33e-24  & 1.28e-23  & 4.68e-23  & 1.32e-22  & 3.89e-22  \\

Ursa Major I               & 1.11e-25  & 1.13e-25  & 1.87e-25  & 3.15e-25  & 5.90e-25  & 1.61e-24  & 3.87e-24  & 1.01e-23  & 3.89e-23  & 1.12e-22  & 3.30e-22  \\

Ursa Major II              & 5.87e-26  & 7.36e-26  & 1.38e-25  & 2.17e-25  & 3.43e-25  & 6.92e-25  & 1.37e-24  & 3.09e-24  & 1.05e-23  & 2.82e-23  & 7.85e-23  \\

Ursa Minor                 & 3.46e-26  & 3.27e-26  & 4.75e-26  & 7.98e-26  & 1.59e-25  & 4.65e-25  & 1.15e-24  & 2.94e-24  & 1.08e-23  & 3.01e-23  & 8.75e-23  \\

Willman 1                  & 1.22e-25  & 1.38e-25  & 2.37e-25  & 3.69e-25  & 6.21e-25  & 1.51e-24  & 3.40e-24  & 8.29e-24  & 2.94e-23  & 8.03e-23  & 2.27e-22  \\

Combined                   & 2.52e-26  & 2.90e-26  & 5.00e-26  & 7.91e-26  & 1.34e-25  & 3.03e-25  & 6.23e-25  & 1.36e-24  & 3.96e-24  & 9.37e-24  & 2.40e-23  \\

\end{tabular}
\end{ruledtabular}
    
\end{table}

\begin{table}[h] \scriptsize
\caption{\label{tab:ww_limits} 95\% CL limits on the annihilation cross section for the \ww channel ($\mathrm{cm}^3\,\second^{-1}$). }

\begin{ruledtabular}
\begin{tabular}{ l  c c c c c c c c}
 & 81\GeV  & 100\GeV  & 250\GeV  & 500\GeV  & 1000\GeV  & 2500\GeV  & 5000\GeV  & 10000\GeV \\
\hline
Bootes I                   & 6.82e-25  & 8.97e-25  & 2.76e-24  & 6.78e-24  & 1.82e-23  & 7.66e-23  & 2.46e-22  & 8.12e-22  \\

Bootes II                  & --  & --  & --  & --  & --  & --  & --  & --  \\

Bootes III                 & --  & --  & --  & --  & --  & --  & --  & --  \\

Canes Venatici I           & 6.61e-24  & 8.38e-24  & 2.56e-23  & 6.60e-23  & 1.86e-22  & 8.17e-22  & 2.62e-21  & 9.00e-21  \\

Canes Venatici II          & 5.43e-24  & 6.06e-24  & 1.11e-23  & 2.27e-23  & 5.56e-23  & 2.28e-22  & 7.71e-22  & 2.90e-21  \\

Canis Major                & --  & --  & --  & --  & --  & --  & --  & --  \\

Carina                     & 1.75e-24  & 1.95e-24  & 5.77e-24  & 1.94e-23  & 6.07e-23  & 2.89e-22  & 1.03e-21  & 4.00e-21  \\

Coma Berenices             & 1.94e-25  & 2.38e-25  & 6.55e-25  & 1.62e-24  & 4.43e-24  & 1.95e-23  & 6.70e-23  & 2.52e-22  \\

Draco                      & 2.79e-25  & 3.44e-25  & 9.71e-25  & 2.36e-24  & 6.11e-24  & 2.38e-23  & 7.37e-23  & 2.53e-22  \\

Fornax                     & 2.25e-24  & 2.77e-24  & 7.35e-24  & 1.70e-23  & 4.25e-23  & 1.67e-22  & 5.30e-22  & 1.85e-21  \\

Hercules                   & 1.11e-23  & 1.33e-23  & 3.17e-23  & 6.78e-23  & 1.58e-22  & 5.36e-22  & 1.43e-21  & 4.07e-21  \\

Leo I                      & 6.51e-24  & 7.50e-24  & 1.66e-23  & 3.69e-23  & 9.47e-23  & 4.05e-22  & 1.39e-21  & 5.29e-21  \\

Leo II                     & 3.73e-24  & 4.27e-24  & 1.03e-23  & 2.53e-23  & 7.19e-23  & 3.43e-22  & 1.25e-21  & 4.95e-21  \\

Leo IV                     & 1.64e-24  & 2.02e-24  & 5.99e-24  & 1.58e-23  & 4.64e-23  & 2.25e-22  & 8.23e-22  & 3.27e-21  \\

Leo V                      & --  & --  & --  & --  & --  & --  & --  & --  \\

Pisces II                  & --  & --  & --  & --  & --  & --  & --  & --  \\

Sagittarius                & --  & --  & --  & --  & --  & --  & --  & --  \\

Sculptor                   & 1.41e-24  & 1.77e-24  & 5.02e-24  & 1.18e-23  & 2.85e-23  & 8.91e-23  & 2.24e-22  & 6.67e-22  \\

Segue 1                    & 3.99e-25  & 4.97e-25  & 1.35e-24  & 3.08e-24  & 7.48e-24  & 2.54e-23  & 6.21e-23  & 1.70e-22  \\

Segue 2                    & --  & --  & --  & --  & --  & --  & --  & --  \\

Sextans                    & 1.65e-24  & 1.78e-24  & 3.44e-24  & 7.44e-24  & 1.91e-23  & 8.20e-23  & 2.84e-22  & 1.09e-21  \\

Ursa Major I               & 6.53e-25  & 8.03e-25  & 2.29e-24  & 5.85e-24  & 1.65e-23  & 7.51e-23  & 2.62e-22  & 9.93e-22  \\

Ursa Major II              & 4.08e-25  & 4.67e-25  & 9.39e-25  & 1.86e-24  & 4.22e-24  & 1.54e-23  & 4.80e-23  & 1.68e-22  \\

Ursa Minor                 & 1.73e-25  & 2.18e-25  & 6.68e-25  & 1.72e-24  & 4.76e-24  & 2.04e-23  & 6.71e-23  & 2.40e-22  \\

Willman 1                  & 7.27e-25  & 8.67e-25  & 2.11e-24  & 4.74e-24  & 1.18e-23  & 4.44e-23  & 1.24e-22  & 3.74e-22  \\

Combined                   & 1.56e-25  & 1.84e-25  & 4.12e-25  & 8.37e-25  & 1.75e-24  & 4.52e-24  & 1.10e-23  & 3.27e-23  \\

\end{tabular}
\end{ruledtabular}
    
\end{table}


\end{turnpage}

\begin{table}[h]
\begin{ruledtabular}
\caption{\label{tab:systematics} 
Systematic uncertainties on the maximum likelihood analysis decomposed into contributions from the IRFs, the diffuse modeling, and the spatial extension of the dark matter profile. 
Entries represent the percentage change to the combined upper limits on the dark matter annihilation cross section from varying each component individually. }

\begin{ruledtabular}
\begin{tabular}{ l c c c c c }
 &  & 10\GeV  & 100\GeV  & 1000\GeV  & 10000\GeV \\
\hline
\hline 
\multirow{3}{*}{$e^{+}e^{-}$}  & IRFs & +14\%/-12\% & +12\%/-10\% & +11\%/-9\% & +11\%/-9\% \\
 & Diffuse &  +3\%/-4\% &  +3\%/-3\% &  +1\%/-1\% &  +1\%/-1\% \\
 & Extension &  +7\%/-5\% & +17\%/-11\% & +11\%/-6\% & +10\%/-6\% \\
\hline 
\multirow{3}{*}{$\mu^{+}\mu^{-}$}  & IRFs & +15\%/-12\% & +12\%/-10\% & +11\%/-9\% & +11\%/-9\% \\
 & Diffuse &  +4\%/-5\% &  +3\%/-4\% &  +2\%/-1\% &  +1\%/-1\% \\
 & Extension &  +7\%/-5\% & +15\%/-10\% & +11\%/-6\% & +10\%/-6\% \\
\hline 
\multirow{3}{*}{$\tau^{+}\tau^{-}$}  & IRFs & +15\%/-13\% & +12\%/-10\% & +13\%/-11\% & +12\%/-10\% \\
 & Diffuse &  +5\%/-5\% &  +1\%/-5\% &  +1\%/-1\% &  +0\%/-1\% \\
 & Extension &  +6\%/-4\% & +14\%/-9\% & +13\%/-7\% &  +6\%/-3\% \\
\hline 
\multirow{3}{*}{$u \bar u$}  & IRFs & +15\%/-14\% & +14\%/-12\% & +12\%/-10\% & +11\%/-9\% \\
 & Diffuse &  +9\%/-4\% &  +3\%/-4\% &  +3\%/-4\% &  +2\%/-3\% \\
 & Extension &  +4\%/-3\% &  +9\%/-7\% & +12\%/-8\% & +12\%/-7\% \\
\hline 
\multirow{3}{*}{$b \bar b$}  & IRFs & +15\%/-13\% & +14\%/-12\% & +12\%/-10\% & +11\%/-9\% \\
 & Diffuse & +10\%/-3\% &  +4\%/-5\% &  +2\%/-4\% &  +2\%/-3\% \\
 & Extension &  +3\%/-2\% &  +8\%/-6\% & +12\%/-8\% & +12\%/-7\% \\
\hline 
\multirow{3}{*}{$W^{+}W^{-}$}  & IRFs & -- & +14\%/-12\% & +12\%/-10\% & +13\%/-11\% \\
 & Diffuse & -- &  +4\%/-4\% &  +2\%/-4\% &  +1\%/-1\% \\
 & Extension & -- &  +8\%/-6\% & +13\%/-8\% & +14\%/-8\% \\

\end{tabular}
\end{ruledtabular}

\end{ruledtabular}
\end{table}



\end{document}